\documentclass[aps,prd,preprint,onecolumn,nofootinbib,superscriptaddress,amsmath,amssymb]{revtex4-2}
\usepackage{setspace}
\usepackage{hyperref}
\usepackage{amsmath, amssymb, amsfonts}
\usepackage{graphicx}
\usepackage{subcaption}
\usepackage{graphicx}
\usepackage{tikz}
\usetikzlibrary{arrows.meta, positioning}
\usepackage{hyperref}
\usepackage{physics}
\usepackage{comment}
\usepackage{multirow}
\usepackage{braket}
\usepackage{tikz}
\usepackage{booktabs}
\usepackage{float}
\usepackage{caption}
\usepackage{booktabs}
\usepackage{multirow}
\usepackage{makecell}
\usepackage{xcolor}
\usepackage{listings}
\usetikzlibrary{shapes,arrows,positioning,calc}

\RequirePackage{mathtools}
\RequirePackage{mathrsfs} %need this for mathscr

\RequirePackage{amsthm}
\RequirePackage{stmaryrd} %for \llbracket

\usepackage{multirow}
\usepackage{youngtab}

\newcommand{\SuSy}{SuSy}
\newcommand{\CoDy}{CoDy}

\newif\ifFSfinal
\FSfinalfalse
\newcommand{\SU}{\mathrm{SU}}

\newcommand{\Sp}{\mathrm{Sp}}
\newcommand{\SO}{\mathrm{SO}}

 \graphicspath{{Figs/}{LatticeFigs/}}

\theoremstyle{plain}

\theoremstyle{definition}

\RequirePackage{tikz}
\RequirePackage{pgfplots}
\pgfplotsset{compat=1.16}
\usetikzlibrary{positioning,automata,calc,fit,shapes,backgrounds,arrows.meta}
\usepackage{makecell}
\usepackage{graphicx}
\usepackage{array} % da mettere nel preambolo
\usepackage{makecell} 

\usepackage{subcaption}
\RequirePackage[utf8]{inputenc}
\usepackage{marvosym} %/Lightning
\usepackage{subcaption} %subfigures
\usepackage{tabularx,booktabs}   % da inserire nel preambolo
\begin{document}

\title{SuSy and CoDy Radiation from Evaporating Black Holes
%in GR and Beyond
%CoDy
}

%%% --- title block, styled after the group's recent RevTeX papers -----------
\author{Giacomo Cacciapaglia}
\affiliation{Laboratoire de Physique Th\'eorique et Hautes \'Energies (LPTHE), UMR 7589, Sorbonne Universit\'e \& CNRS, 75252 Paris, France}
\affiliation{{$\hbar$QTC, Quantum Theory Center,   University of Southern Denmark, Campusvej 55, DK-5230 Odense, Denmark}}

\author{Francesco Sannino}
\affiliation{{$\hbar$QTC, Quantum Theory Center,   University of Southern Denmark, Campusvej 55, DK-5230 Odense, Denmark}}
\affiliation{\mbox{Dip. di Fisica ``E.\ Pancini'', Universit\`a di Napoli Federico II, Via Cintia, 80126 Napoli, Italy}}
\affiliation{Institute for Particle Physics Phenomenology (IPPP), Durham Univ., South Road, DH1 3LE, Durham, United Kingdom}

\author{Paola Simone}
\affiliation{Scuola Superiore Meridionale, Universit\`a di Napoli Federico II, Largo San Marcellino 10, 80138 Napoli, Italy}

\author{Vania Vellucci}
\affiliation{{$\hbar$QTC, Quantum Theory Center,   University of Southern Denmark, Campusvej 55, DK-5230 Odense, Denmark}}

\begin{abstract}
\setstretch{1}
We investigate the interplay between physics beyond the Standard Model
and modified spacetime geometries in shaping the photon and neutrino
spectra from evaporating black holes. We implement three representative
supersymmetric (SuSy) benchmarks and a composite dynamics (CoDy)
scenario in \texttt{BlackHawk}. % allowing us to study their impact on Hawking evaporation and the resulting particle spectra. 
We consider
both Schwarzschild black holes and regular black holes geometries, namely
polymerized, Hayward, and Simpson--Visser black holes. New degrees
of freedom at the TeV scale become accessible at Hawking temperatures
corresponding to black hole masses of approximately
$M_{\rm BH}\sim 10^{10}\,\mathrm{g}$. We show that physics beyond the Standard Model produces distinctive
signatures in both the instantaneous secondary-photon and neutrino
spectra. Radiative decays generate narrow spectral features with
approximately temperature-independent relative amplitudes, while
multistep decay cascades produce broad enhancements that become
increasingly pronounced at higher temperatures. In contrast,
compressed mass spectra exhibit no localized features. Crucially, we show that the effects of new particle physics and
modified spacetime geometries can become intertwined in specific
energy regimes, altering the resulting spectral signatures.
Our results demonstrate that photon and neutrino spectra provide
complementary probes of new particle physics and black hole geometry,
with predicted spectral enhancements extending into the energy
ranges accessible to current and future high-energy observatories.
\end{abstract}
 
\maketitle

\clearpage

\tableofcontents

\newpage

%====================================================================================
\section{Introduction}

The Standard Model (SM) of particle physics~\cite{Glashow:1961tr,Weinberg:1967tq,Salam:1968rm} and Einstein's theory of General Relativity (GR)~\cite{Einstein:1915ca,Einstein:1916vd} are, to date, the most successful theories describing particle interactions and gravitation. Notwithstanding their theoretical and experimental successes, there are multiple reasons to believe that both should be at least extended or modified. In the SM, an important theoretical issue arises in the electroweak (EW) sector, specifically related to the Brout-Englert-Higgs mechanism~\cite{Englert:1964et,Higgs:1964pj,Guralnik:1964eu}: the so-called Naturalness problem~\cite{tHooft:1979rat}. Setting to zero the Higgs mass parameters does not increase the symmetries of the SM, hence radiative loop corrections are proportional to any new scale that couples to the Higgs field. As a consequence, the Higgs mass should naturally be close to the largest physical scale, such as the Planck one $M_\text{P} \sim 1.22\cdot 10^{16}$~GeV, and an extreme and ``un-natural'' fine-tuning would be needed to explain why the Higgs boson is so light ($\sim 125$~GeV)~\cite{ATLAS:2012yve,CMS:2012qbp}. During the last demicentury, considerable effort has been spent in developing models whose Higgs sector is natural, typically by extending some of the symmetries of the SM to larger groups and giving rise to a plethora of Beyond the Standard Model (BSM) scenarios. Two major realizations stand out: Supersymmetry (\SuSy)~\cite{Witten:1981nf,Sakai:1981gr} and Composite Dynamics (\CoDy)~\cite{Weinberg:1975gm}. In the \SuSy\ scenario, models are built upon the only possible extension of spacetime symmetries, i.e. the Poincaré algebra, where the new symmetry generators relate particles of different spin. Instead, the \CoDy\ scenario consists in extending the gauge symmetry of the SM by a new confining force, whose charged matter fermions replace the Brout-Englert-Higgs sector. In both cases, the main theoretical and phenomenological result is the prediction of new heavy particles and interactions around the TeV scale~\cite{Barbieri:2000gf}. Models based on \SuSy\ and \CoDy\ are currently being tested in collider searches~\cite{ParticleDataGroup:2026mpi}, without yet having statistical evidence for the existence of new particles due to the limited energy range of colliders and to the difficulty to distinguish a clean signature of a new heavy particle from the SM backgrounds. Notably, the SM also falls short of describing well-established phenomena, such as the asymmetry of matter over antimatter, the presence of dark matter and dark energy, neutrino masses and cosmic inflation: many BSM models addressing naturalness incorporate at least one of them.

In GR, instead, the unavoidable formation of spacetime singularities following gravitational collapse~\cite{Penrose:1964wq,Hawking:1970zqf} represents a breakdown of the predictability of the GR theory, and highlights its incompleteness. Such singularities are often expected to be ``hidden'' behind a Black Hole (BH) horizon~\cite{Penrose:1964wq}.
The singularity issue in GR can be addressed via a violation of the strong energy condition in the vicinity of the centre of BHs~\cite{Zaslavskii:2010qz}, leading to non-singular spacetimes~\cite{Hayward:2005gi,Simpson:2018tsi}, known as regular black holes (RBHs). These are thought as possible solutions of a more complete, potentially quantum, theory of gravity that is expected to modify GR dynamics at lengths smaller than a characteristic scale $\ell$, effectively resolving the central singularity. Distinguishing the metric of RBHs from that of the singular BHs, stemming from GR, could be possible in principle through the observation of gravitational waves signal coming from their coalescence \cite{Cardoso:2016rao,Franzin:2023slm,Carballo-Rubio:2025fnc,Coviello:2025pla}, however this usually requires compact object mergers of masses well below the ones of astrophysical BHs and an experimental sensitivity far beyond the current one.

In this work, we aim at investigating whether BSM scenarios and RBH metrics can affect the physics of BHs in an observable way. The only physical process where such effects may be observed is the emission of particles via the semiclassical process known as Hawking radiation~\cite{Hawking:blackholeexplosions, Hawking:particlecreation}. BHs, in fact, have a quasi-thermal nature and can evaporate by radiating away particles with a spectrum analogue to the blackbody radiation. As long as the BH is light enough that its Hawking temperature overcomes the appropriate mass scale, any particles existing in nature will be produced. In particular, a Hawking temperature of a few TeV, where the effects of \SuSy\ and \CoDy\ models is expected, corresponds to BH masses around $10^{10}$~g, with a radius of the order of $10^{-4}$~fermi ($10^{-19}$~metres). 
Henceforth, the presence of new particles around the TeV scale would modify the spectrum of BHs lighter than about $10^{10}$~g and shorten their remaining evaporation time. Some \SuSy\ scenarios have already been considered in the literature~\cite{Ukwatta:2015iba,Baker:2022rkn}, in this work we analyse specific benchmark models based on the minimal (MSSM) and next-to-minimal (NMSSM) realizations, which are compatible with the Higgs physics and the current collider bounds. Furthermore, we consider \CoDy\ models where the Higgs boson emerges as a light, pion-like, bound state~\cite{Kaplan:1983fs} of a QCD-like confining dynamics. In this case, the BHs radiate the constituent fermions and gauge bosons, which then hadronise in a similar fashion as quarks do in the SM. We demonstrate how each benchmark affects the spectrum of emitted photons and neutrinos, and provide numerical tables that can be used to model the detailed time-dependent spectra. For completeness, note that other new physics cases have been considered in the literature under various conditions, including dark matter~\cite{Allahverdi:2017sks,Lennon:2017tqq,Gondolo:2020uqv,Cheek:2021odj,Cheek:2021cfe,Gehrman:2023esa}, axions and axion-like particles~\cite{Bernal:2021yyb,Agashe:2022phd,Jho:2022wxd}, mirror and dark sectors~\cite{Bell:1998jk,Baker:2021btk}, heavy neutrinos~\cite{Lunardini:2019zob,DeRomeri:2024zqs}. 
The shape of the spectra are also affected by modifications of the BH metric, such as that provided by RBHs~\cite{Arbey:2026koc}. For RBHs, an appropriate value of the scale $\ell$ could give rise to effects comparable to that of TeV-scale BSM models. Hence, deviations in the Hawking radiation spectrum can probe both model of new physics and the quantum nature of gravity. 

In recent years, several numerical tools have been developed to simulate BH evaporation. \texttt{BlackHawk} ~\cite{Arbey:2019mbc,Arbey:2021mbl} is an open-source code that allows users to compute both primary and secondary spectra of emitted particles for a given BH mass, spin, and, optionally, mass and spin distributions. Although the code allows for the inclusion of additional BSM particles, its standard pipeline for computing secondary spectra is primarily designed for SM states. In this work, we incorporated our BSM numerical tables into the \texttt{BlackHawk} framework and extended the existing pipeline to account for the emission and subsequent evolution of the new particle species. RBH metrics, instead, were recently incorporated in \texttt{BlackHawk v3.0}\cite{Arbey:2026koc}, which we use for our numerical studies.

At the observational level, the most likely source of Hawking radiation in the present universe is offered by primordial black hole (PBH) survivors, whose final evaporation could be happening now in the vicinity of the solar system. Their signature can be targeted by gamma-ray and neutrino telescopes, as BH  final evaporation is expected to produce characteristic high-energy signatures that could distinguish PBH bursts from conventional astrophysical transients \cite{Ukwatta:2015iba}. Searches performed by Fermi-LAT, HAWC, H.E.S.S., and LHAASO have placed increasingly stringent constraints on the local PBH burst rate \cite{FermiLAT:2018PBH,HAWC:2020PBH,HESS:2023PBH,LHAASO:2025PBH}. In parallel, phenomenological studies have investigated the potential of current and future gamma-ray detectors to probe BSM dark sectors through the observation of nearby PBH bursts \cite{Baker:2025ffi}. Recently, a new source was proposed via the possible production of small black holes, dubbed ``black morsels'', during the merger of astrophysical BHs~\cite{Cacciapaglia:2024wtp} which have been discovered via their gravitational wave emission. This process could be allowed in some cases within GR or via extensions~\cite{Cacciapaglia:2026jea}, and a preliminary search using Fermi-LAT data is under way~\cite{Acharyya:2025yzx}.

The paper is organised as follows: after presenting the general methodology used in this work in Section~\ref{sec:methodology}, we discuss BSM benchmarks and their impact on the Hawking spectra in Section~\ref{sec:BSM}. In Section~\ref{sec:RBHs} we repeat the analysis for RBH metrics, comparing the spectra with those of BSM benchmarks. In Section~\ref{sec:detection} we briefly characterise how the modified photon spectra could affect the observations at gamma-ray telescopes, before offering our conclusions and outlook in Section~\ref{sec:concl}. More details on the spectra are provided in the appendix~\ref{sec:appendix}.

%=====================================================================================
%\section{Stable particles spectra for an evaporating black hole}
\section{Methodology}\label{sec:methodology}
Hawking radiation~\cite{Hawking:blackholeexplosions,Hawking:particlecreation} consists of the emission of elementary particles, following a semiclassical tunnelling past the gravitational barrier of the BH. All particles with a mass below a characteristic energy scale, the Hawking temperature, can be emitted. This temperature, which can be thought of as the thermodynamic characteristic of the spectrum as measured at infinite distance from the horizon, is inversely proportional to the BH mass so that the smaller the BH the hotter the emitted radiation.
The primary Hawking flux, corresponding to particles emitted directly by the BH, is given for a particle species \(i\) with energy \(E\) and a BH of mass \(M\) by:
\begin{equation}
\frac{d^2 N_i}{dt\, dE}
=
g_i\cdot \sum_{\ell,\,s}
\frac{\Gamma_{\ell,s}(E,M)}{2\pi}
\frac{1}{\exp\!\left(\frac{E}{T_H}\right)\pm 1}\,,
\end{equation}
where $g_i$ is the number of degrees of freedom of the particle and $T_H$ the Hawking temperature. The $\Gamma _{ls}$ function, dubbed ``greybody'' factor~\cite{Page:1976df}, quantifies the tunnelling probability: it is here that information about both the black hole metric and the particle properties (i,e, its spin) are encoded. The primary particles, however, cannot reach directly an observer, as they first undergo processes such as radiation, decays and, in the case of QCD charged states, hadronisation. These physical phenomena produce spectra of secondary particles, consisting of only the stable particles in the theory, which are connected to the primary spectra via specific transfer functions:
\begin{equation}
      \frac{d^2 N_f}{dt\, dE}  = \sum_i \int dE'\ \mathcal{F}_{i\to f} (E',E)\ \frac{d^2 N_i}{dt\, dE'}\,,
\end{equation}
for a secondary particle $f$ of energy $E$.
In the SM, secondary particle consist of neutrinos, protons  and antiprotons, electrons and positrons, and photons. In some BSM scenarios, other stable particles may emerge, such as dark matter candidates and cosmologically long-lived axions. The final observable spectrum, therefore, will crucially depend on the physical spectrum of the particle theory, even at energies not reachable at colliders~\cite{Ukwatta:2015iba}. The greybody factors only depend on the spin of the primary particles, nevertheless they are affected by modification of GR, such has RBH metrics~\cite{Arbey:2026koc}. For these reasons, the final spectrum of stable particles deriving from a radiating BH can in principle provide insights both into the physics of BSM, but also on deviations from classical GR.

Our goal is to assess the impact of our benchmark BSM models and RBH metrics on the secondary spectra, for which we follow the strategy employed in the public code  \texttt{BlackHawk}~\cite{arbey2019blackhawk,Arbey:2021mbl}. The code computes separately the greybody factors of a BH with specific mass and spin, while the transfer functions are archived in the form of numerical tables for each pair of initial and final particles. The effect of RBH metrics on the greybody factors has been recently included in the \texttt{BlackHawk v3.0} release~\cite{Arbey:2026koc}, which we will use in the following. Our main scope is, therefore, to produce transfer function tables for the \SuSy\ and \CoDy\ benchmarks.

\begin{figure}[h!]
    \centering

    % First row
    \begin{subfigure}[t]{0.31\textwidth}
        \centering
        \includegraphics[width=\linewidth]{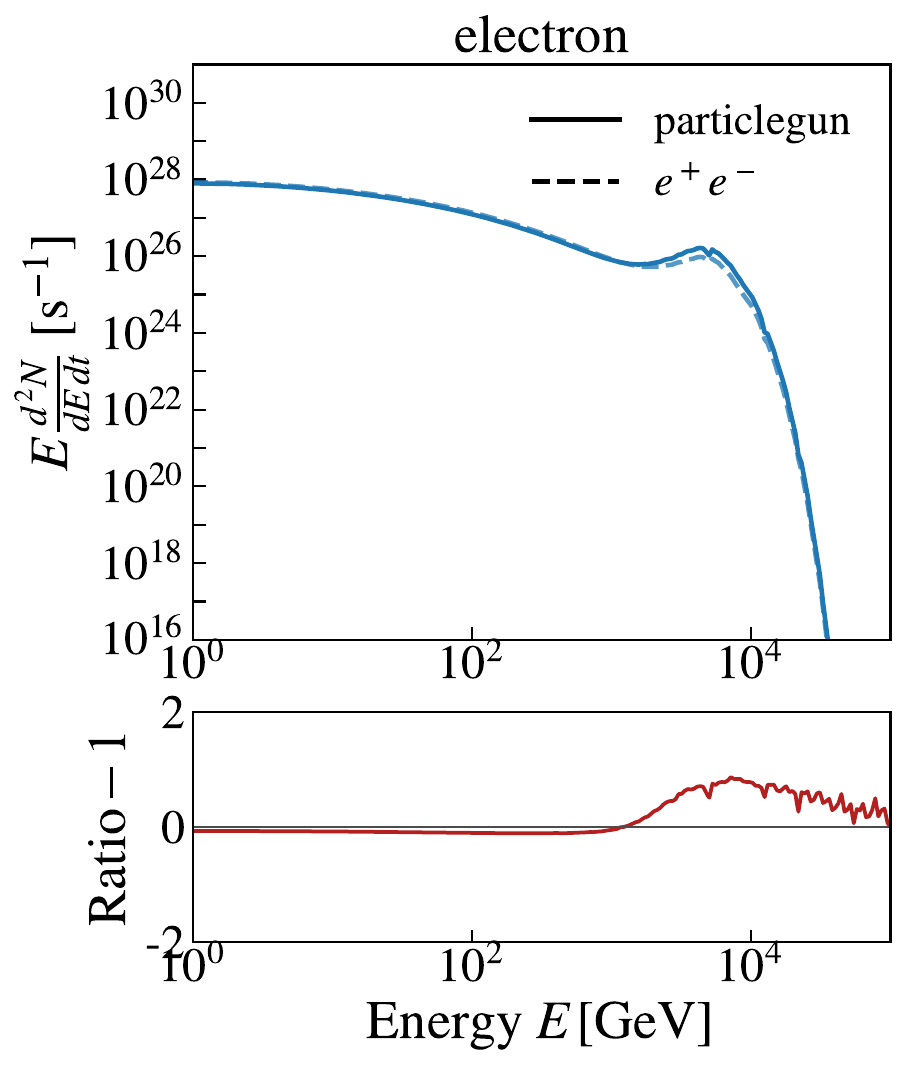}
        \caption{}
        \label{fig:fig1}
    \end{subfigure}
    \hfill
    \begin{subfigure}[t]{0.31\textwidth}
        \centering
        \includegraphics[width=\linewidth]{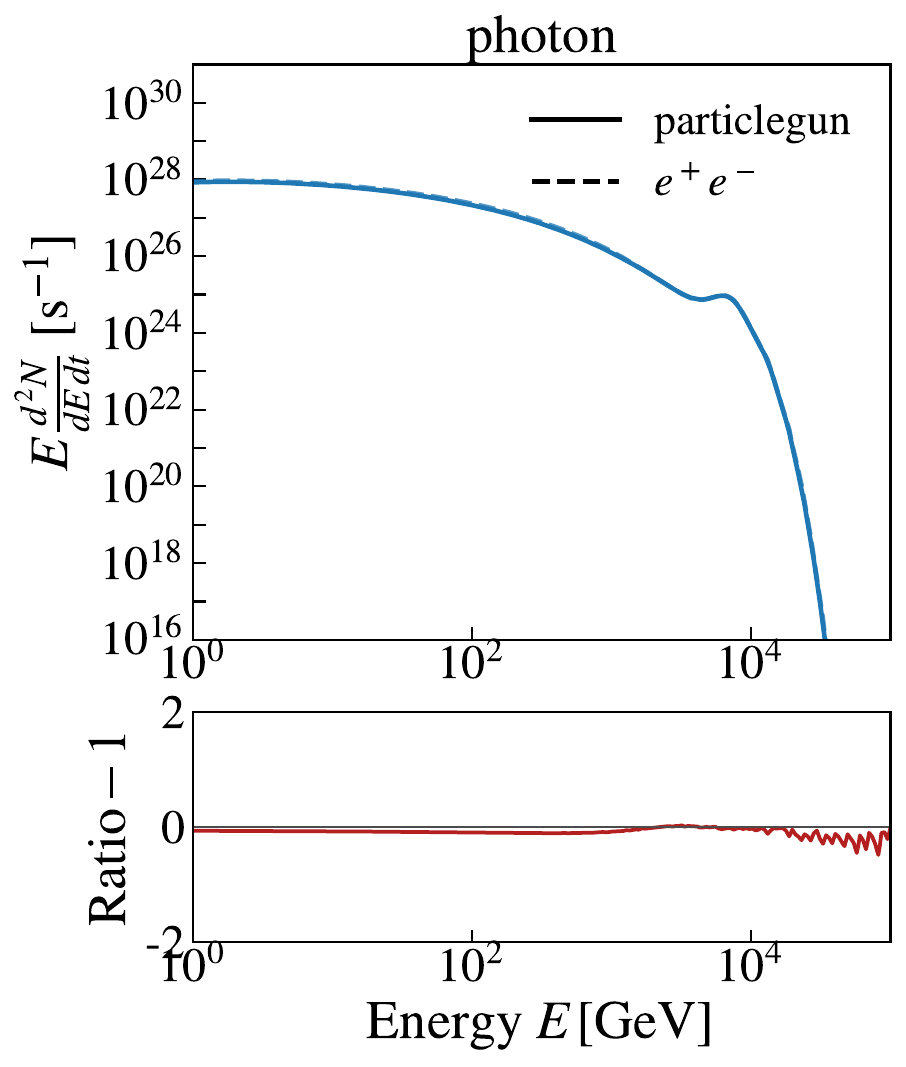}
        \caption{}
        \label{fig:fig2}
    \end{subfigure}
    \hfill
    \begin{subfigure}[t]{0.31\textwidth}
        \centering
        \includegraphics[width=\linewidth]{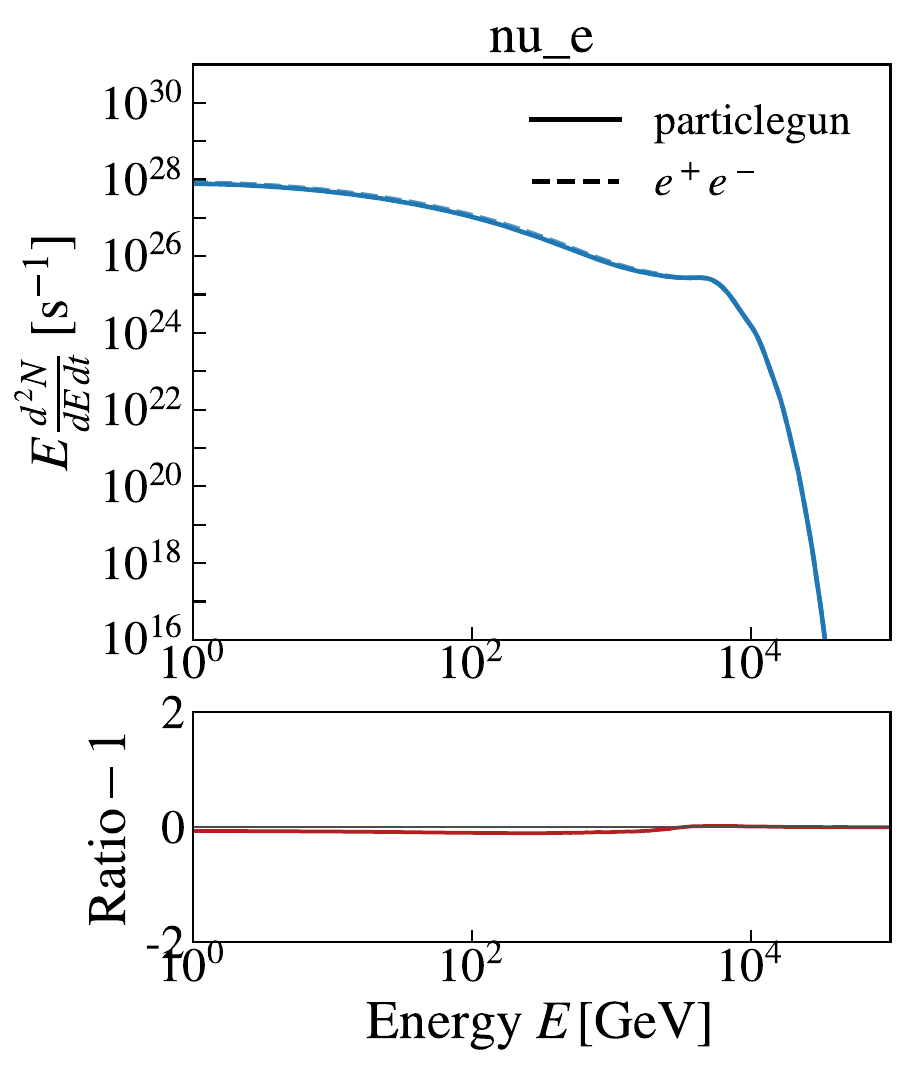}
        \caption{}
        \label{fig:fig3}
    \end{subfigure}

    \vspace{0.3cm}

    % Second row
    \begin{subfigure}[t]{0.31\textwidth}
        \centering
        \includegraphics[width=\linewidth]{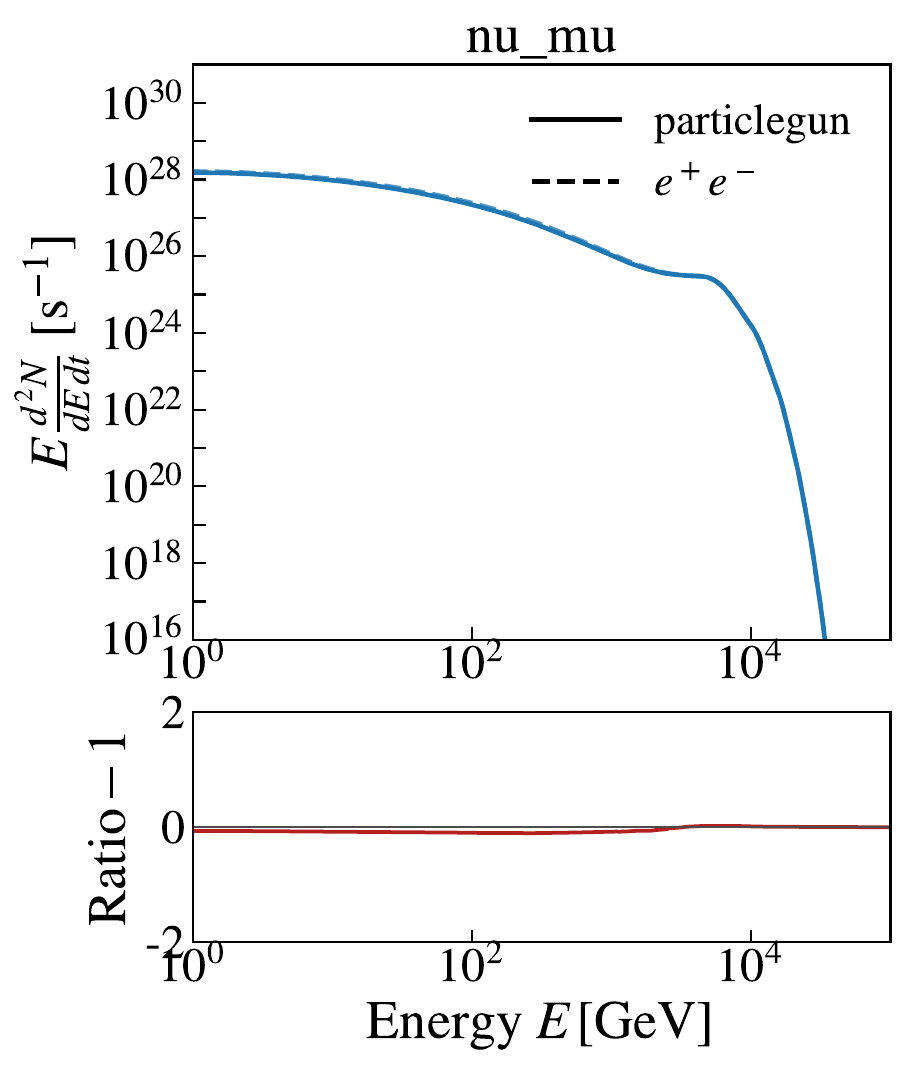}
        \caption{}
        \label{fig:fig4}
    \end{subfigure}
    \hfill
    \begin{subfigure}[t]{0.31\textwidth}
        \centering
        \includegraphics[width=\linewidth]{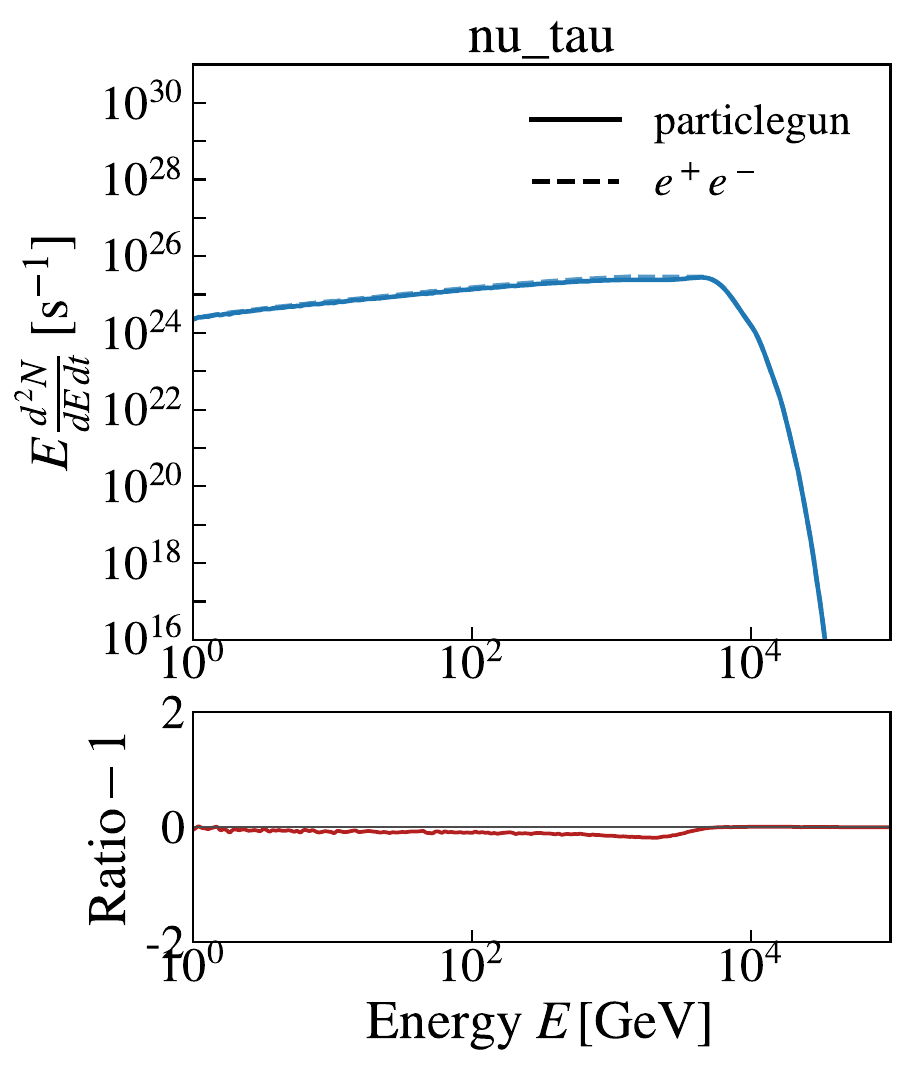}
        \caption{}
        \label{fig:fig5}
    \end{subfigure}
    \hfill
    \begin{subfigure}[t]{0.31\textwidth}
        \centering
        \includegraphics[width=\linewidth]{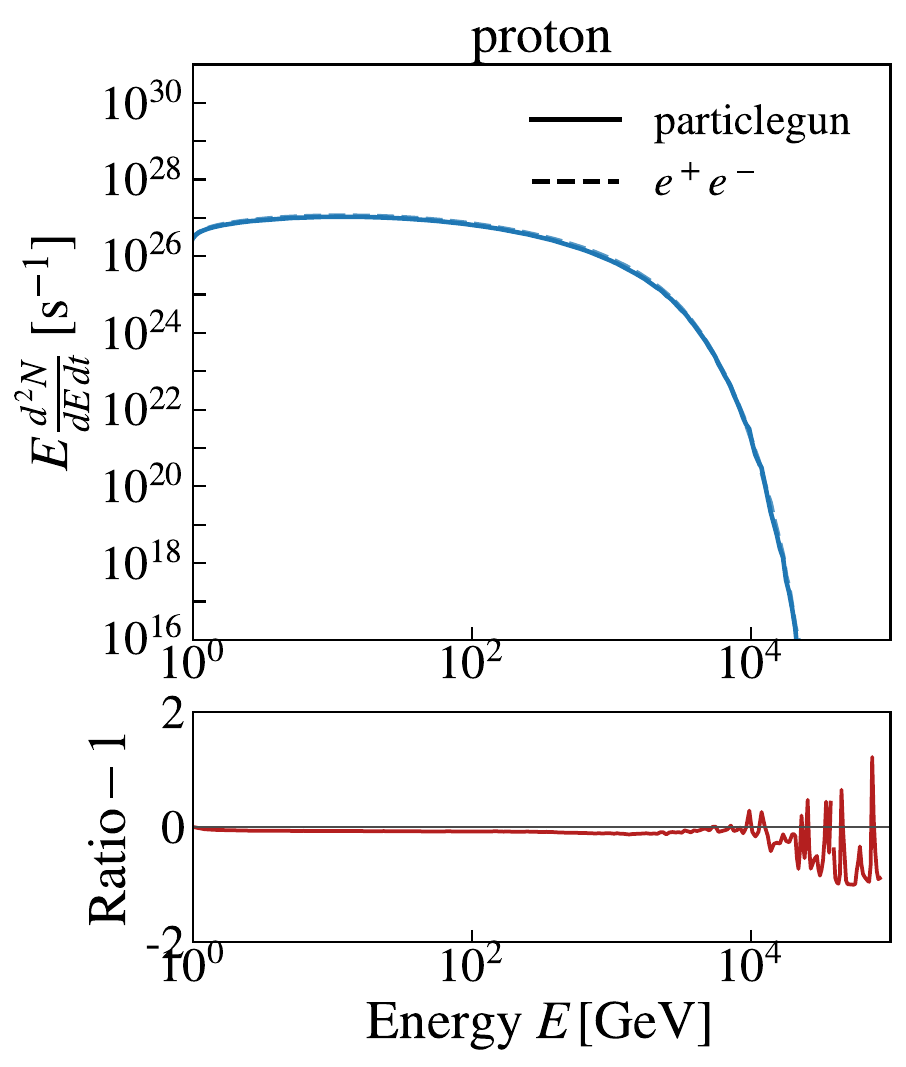}
        \caption{}
        \label{fig:fig6}
    \end{subfigure}

    \caption{Comparison between the stable secondary spectra of the SM produced using \texttt{BlackHawk v3.0} with $e^+e^-$ collisions and the recomputed tables using the particle gun method. In the subplot, we show the ratio between the spectra derived with \texttt{BlackHawk v3.0} and the one with our method. }
    \label{fig:comparison_SM}
\end{figure}

For the SM~\cite{Arbey:2019mbc}, and few examples of BSM~\cite{Arbey:2021mbl}, \texttt{BlackHawk} uses \texttt{Pythia}~\cite{sjostrand2015introduction}, a code developed within the context of collider physics. The main strategy consists in producing a pair of two primary particles via a fictitious electron-positron collision at fixed energy $2E$, then process the event to compute the spectrum of particles observed at large distances. For particles that cannot be directly produced, such as the Higgs boson, a fictitious resonance $X$ is included which decays into the desired primary particle. This method cannot be directly transported to any BSM model: for instance, \SuSy\ models where only implemented in \texttt{Pythia} specifically for proton collisions. To circumvent this limitation, we employ the standard ``particle gun'' set up in \texttt{Pythia}, which allows to feed a single particle of energy $E$ in the event record and process its evolution via radiation, decays and hadronisation, without relying on any production method. Additionally, this allows to avoid the problem of off-shell production of resonances, as also pointed out in~\cite{Wu:2025ovd}. This method is however ineffective for primary particles charged under QCD, as the final state after hadronisation can only be colour-neutral. Hence, to study coloured primaries we must define a fictitious resonance that couples to all coloured particles in the model, via the process
 \begin{equation}
    \mathrm{X \rightarrow q_i \bar{q_i} \rightarrow QCD-hadronization}\,.
    \label{hadro}
 \end{equation}
To avoid off-shell production, we redefine the rest mass for every re-initialisation to be $M_X = E_\text{init}$, where $E_\text{init}$ is the initial energy of the coloured particle under study.
As a consistency check, we recomputed the SM tables with this particle-gun method and verified that it gives results compatible with the tables already contained in \texttt{BlackHawk v3.0}. In Figure~\ref{fig:comparison_SM} we show the secondary spectra for photons obtained with the original \texttt{BlackHawk v3.0} tables and the ones we recomputed. As a final remark, we recall that \texttt{Pythia} can handle primary particle energies between $5$~GeV to about $100$~TeV, which is compatible with the scope of this work. For higher energies, one would need a specific treatment of radiation and decays, along the line of the work done for the SM~\cite{bauer2021dark}.

As generic \SuSy\ particles are already implemented in \texttt{Pythia} \cite{sjostrand2015introduction}, we simply used the latest version \texttt{Pythia8.3} \cite{Bierlich:2022pfr} to generate all the numerical transfer function tables for the benchmark models under consideration. The supersymmetric particles, in fact, would simply decay into SM ones, with the addition of one stable neutral state, which is often associated to dark matter.

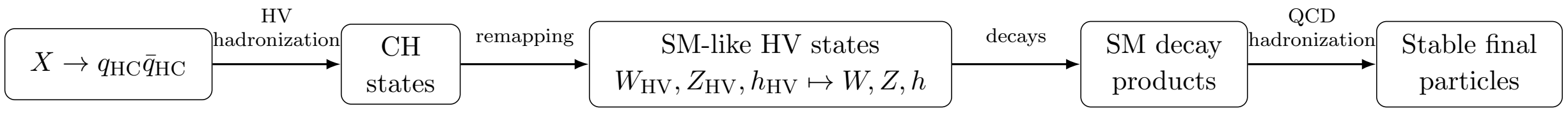
\begin{figure}[h!]
\centering

\resizebox{0.98\textwidth}{!}{%
\begin{tikzpicture}[
  node distance=1.8cm,
  box/.style={
    draw,
    rounded corners,
    align=center,
    minimum height=1cm,
    inner xsep=10pt,
    inner ysep=5pt,
    font=\small
  },
  arrow/.style={
    -{Latex[length=2mm]},
    thick
  },
  process/.style={
    midway,
    above=3pt,
    align=center,
    font=\scriptsize
  }
]

\node[box] (initial)
{$X\to q_{\rm HC}\bar q_{\rm HC}$};

\node[box, right=of initial] (composite)
{CH\\states};

\node[box, right=of composite] (remap)
{SM-like HV states\\
$W_{\rm HV},Z_{\rm HV},h_{\rm HV}
 \mapsto W,Z,h$};

\node[box, right=of remap] (sm)
{SM decay\\products};

\node[box, right=of sm] (final)
{Stable final\\particles};

\draw[arrow] (initial) --
node[process] {HV\\hadronization}
(composite);

\draw[arrow] (composite) --
node[process] {remapping}
(remap);

\draw[arrow] (remap) --
node[process] {decays}
(sm);

\draw[arrow] (sm) --
node[process] {QCD\\hadronization}
(final);

\end{tikzpicture}%
}

\caption{Event-generation flow used for the Hidden Valley
hyperquark particle-gun tables.}
\label{fig:HV}
\end{figure}

\CoDy\ scenarios, instead, are qualitatively different, as they feature a new gauge sector which confines. As the resulting bound states have a finite spatial extension, we should take into account possible tidal suppression of their emission beyond the standard point-like approximation~\cite{Johnson:2018gjr}. Nevertheless, we verified that neglecting these finite-size corrections is a reasonable approximation for the final photon and neutrino spectra considered in this work and decided to keep the greybody factors in the point-like approximation already tabulated in \texttt{BlackHawk v3.0} for our results.
We consider the production of new \CoDy\ states in two steps: 1) if the Hawking temperature is lower than the \CoDy\ scale, we only include the production of SM particles; 2) when $T_H$ is larger than such scale,  we consider production of the \CoDy\ constituents together with SM states. Hence, we need to add a layer which takes into account the hadronisation of the \CoDy\ constituents under the new strong dynamics. For a generic gauge theory, a first-principles implementation of the hadronisation process is not available, as the associated non-perturbative properties are typically extracted from data. For this reason, we choose a Composite Higgs model based on a confining $\mathrm{SU}(3)_{\rm CH}$ gauge with fermions in the fundamental representation, which should behave closely to QCD.  In practice, we treat the \CoDy\ sector via the Hidden Valley (HV) package in \texttt{Pythia8.3}, employing the implementation used in \cite{Cacciapaglia:2025pqo}. The HV module provides running gauge couplings,
final-state showering, and Lund-like string fragmentation for a hidden confining
gauge group. The procedure follows the same general strategy described in
Eq.~(\ref{hadro}), with the following additional steps. After the fictitious CH-colour singlet
resonance X is injected and decayed into \CoDy\
partons, identified as $q_{\rm CH}$ and $g_{\rm CH}$, they are passed to the \texttt{Pythia} HV module, which performs the showering and hadronization into composite states. These states are then mapped into a concrete model, where we manually implement their decays into SM particles following the chosen benchmark. The SM decay products, then, undergo the usual procedure of SM decays and QCD hadronization, performed with \texttt{Pythia}. A schematic description of this workflow is presented in Figure~\ref{fig:HV}.

%=====================================================================================

\section{BSM scenarios for black hole evaporation}\label{sec:BSM}

%Supersymmetry was originally introduced to address the hierarchy problem stemming from quantum corrections to the Higgs mass. The framework enlarges the Poincaré group to a larger symmetry group which predicts a particle content organized into supermultiplets of boson-fermion superpartners. For phenomenological studies, the most explored SuSy extension is known as the Minimal Supersymmetric Standard Model (MSSM) \cite{martin1998supersymmetry}. Another explored scenario is its Next-to-Minimal extension (NMSSM) \cite{Ellwanger:2009dp}, with the addition of a single super-scalar field w.r.t. the MSSM. Over the last decades, much experimental effort has been devoted to the search of collider signatures of these superpartners, leading to lower bounds on the SuSy scale. Additionally, SuSy has further provided viable dark matter candidates, and it is compatible with gauge coupling unification at high energies. For these reasons, the theory remains of considerable scientific interest, and either its confirmation or its falsification would represent a major step forward in our understanding of fundamental particle interactions. 

Supersymmetric extensions of the SM provide a theoretically well-motivated framework for addressing several open questions in particle physics and cosmology. \SuSy\ relates bosonic and fermionic degrees of freedom by introducing a superpartner for each SM particle, with spin differing by one half~\cite{Wess:1974tw}: a complex scalar for each fermion and a fermion for each gauge boson. The minimal realization, the Minimal Supersymmetric Standard Model (MSSM), extends the SM particle content with superpartners and requires two Higgs doublets, $H_u$ and $H_d$, to ensure anomaly cancellation and generate masses for up- and down-type fermions. \SuSy\ must be broken at low energies, and its phenomenology is conventionally parametrized by soft supersymmetry-breaking terms~\cite{Iliopoulos:1974zv}, whose structure determines the superpartner spectrum and the resulting experimental signatures \cite{martin1998supersymmetry}.

One of the principal motivations for weak-scale \SuSy\ is the electroweak hierarchy problem~\cite{Dimopoulos:1981zb}. In the SM, the Higgs mass parameter receives radiative corrections sensitive to the scale of new physics. In \SuSy\ theories, contributions from bosonic and fermionic superpartners cancel the leading quadratic divergences, stabilizing the EW scale against large ultraviolet corrections. In particular, the dominant radiative contribution associated with the top-quark Yukawa coupling is compensated by that of the scalar top partners (stops). This cancellation motivates relatively light stops and Higgsinos in scenarios with limited electroweak fine-tuning, although the quantitative degree of naturalness depends on the assumed spectrum and high-scale boundary conditions \cite{Feng:2013pwa}. The observed Higgs-boson mass, $m_h\simeq125$ GeV, provides an important additional constraint: in the MSSM, where the tree-level Higgs mass is bounded by the $Z$-boson mass, substantial radiative corrections are generally required, often favouring heavy stops or significant stop mixing \cite{Draper:2016nbo}.

A second major motivation is the possibility of explaining the cosmological dark matter abundance. If $R$-parity, defined as $R_p=(-1)^{3(B-L)+2s}$, is conserved, the lightest supersymmetric particle (LSP) is stable. In many MSSM scenarios, the LSP is the lightest neutralino, a mixture of the superpartners of the neutral electroweak gauge and Higgs bosons. A neutralino with suitable interactions and mass can reproduce the observed dark matter relic abundance through thermal freeze-out, while also providing targets for direct- and indirect-detection experiments. Alternative candidates, including the gravitino, arise in other \SuSy-breaking frameworks. These possibilities connect collider searches with cosmological observations and underground dark matter experiments \cite{Ellis:2010kf}.

\SuSy\ phenomenology has consequently developed along several complementary directions. Collider studies at the Large Hadron Collider (LHC) investigate the production of coloured superpartners, particularly gluinos and squarks, as well as electroweakinos and sleptons. Searches target final states containing jets, charged leptons, photons and missing transverse momentum, while complementary analyses address $R$-parity violation, compressed mass spectra and long-lived particles, see e.g.~\cite{ATLAS:2024susy}. In parallel, precision calculations of the Higgs mass, Higgs-sector measurements, flavour observables and electric dipole moments constrain the allowed parameter space. Dark matter studies combine relic-density calculations with direct- and indirect-detection limits. Although extensive LHC searches have not established evidence for supersymmetry, their results have significantly constrained many conventional benchmark scenarios, motivating continued investigations of less constrained spectra and non-minimal constructions, such as the Next-to-Minimal Supersymmetric Standard Model (NMSSM) \cite{Ellwanger:2009dp}. \SuSy\ therefore remains a useful framework in which to explore the interplay between electroweak naturalness, Higgs physics, collider phenomenology and the nature of dark matter.

The \CoDy\ approach to a natural Higgs sector is to replace it with a fundamental gauge
dynamics of new fermions, whose chiral symmetry forbids power-law sensitivity to a higher
cutoff. The oldest incarnation is Technicolor (TC) \cite{Weinberg:1975gm,Susskind:1978ms},
with later variations on the theme \cite{Kaplan:1983fs,Kaplan:1983sm}: the EW symmetry
breaks through the new strong dynamics and the Higgs is identified with the lightest scalar
excitation of the fermion condensate. TC alone cannot feed mass to the SM fermions, so an
extended sector (ETC) \cite{Eichten:1979ah} must be introduced; this sector is not a spectator,
as it reshapes the dynamics \cite{Fukano:2010yv,Rantaharju:2017eej,Rantaharju:2019nmh} and
typically lowers the Technicolor-Higgs mass \cite{Foadi:2012bb}.

The combined TC$+$ETC dynamics can be near-conformal, emerging close to the flavour-driven
quantum phase transition between an IR fixed point and a chirally broken phase
\cite{Miransky:1996pd}. Scenarios for this transition range from a BKT-like one
\cite{Kosterlitz:1974sm,Miransky:1984ef,Miransky:1996pd,Holdom:1988gs,Holdom:1988gr,Cohen:1988sq,Appelquist:1996dq,Gies:2005as}
to a jumping transition \cite{Sannino:2012wy}. The smooth case is what we call \emph{walking}
\cite{Holdom:1988gs,Holdom:1988gr}, expected to enhance bilinear fermion operators and so help
generate SM fermion masses. Because the flavour number is an integer, the gauge-fermion
transition cannot be approached continuously; the ETC interactions remedy this by mapping the
problem onto a gauged NJL model with a continuously varied four-fermion coupling, leading, if
the transition is continuous, to \emph{ideal walking}
\cite{Fukano:2010yv,Yamawaki:1996vr,Rantaharju:2017eej,Rantaharju:2019nmh}. The discovery that
higher representations can be (near) conformal \cite{Sannino:2004qp} led to   the
conformal-window phase diagram \cite{Dietrich:2006cm} that now guides lattice studies
\cite{Pica:2017gcb}.

Alternatively, the new dynamics may leave the electroweak symmetry intact while breaking a
global symmetry, so that the Higgs is one of the resulting Goldstone bosons \cite{Kaplan:1983fs}.
Of particular relevance for a walking transition is the interplay between a dilaton-like state
and a pNGB Higgs, possibly enforced by approximate conformal invariance
\cite{Leung:1985sn,Bardeen:1985sm,Yamawaki:1985zg,Sannino:1999qe,Hong:2004td,Dietrich:2005jn,Appelquist:2010gy};
the subject has received renewed interest
\cite{Hong:2004td,Dietrich:2005jn,Goldberger:2008zz,Appelquist:2010gy,Hashimoto:2010nw,Matsuzaki:2013eva,Golterman:2016lsd,Hansen:2016fri,Golterman:2018mfm}
following lattice evidence for a light singlet scalar
\cite{Appelquist:2016viq,Appelquist:2018yqe,Aoki:2014oha,Aoki:2016wnc,Fodor:2012ty,Fodor:2017nlp,Fodor:2019vmw}.
The pattern of chiral symmetry breaking depends on the underlying dynamics
\cite{Peskin:1980gc,Preskill:1980mz,Kosower:1984aw}, and the symmetry may not break at all if
large-distance conformality develops, as discussed for two-index representations
\cite{Sannino:2004qp}, in the universal $\SU(N)$ classification \cite{Dietrich:2006cm}, for
orthogonal and symplectic groups \cite{Sannino:2009aw}, and for exceptional groups
\cite{Mojaza:2012zd}; symmetry-breaking patterns relevant to a pGB Higgs are classified in
\cite{Mrazek:2011iu,Bellazzini:2014yua}.

For one Dirac species the largest non-abelian flavour symmetry is $\SU(2N_f)$ for (pseudo) real
representations and $\SU(N_f)\times\SU(N_f)$ for complex ones. For real and pseudo-real representations we have two patterns \cite{Kosower:1984aw} of chiral symmetry breaking:
for real ones, $\SU(2N_f)\to\SO(2N_f)$, with the condensate in the symmetric two-index
representation;
for pseudo-real ones, $\SU(2N_f)\to\Sp(2N_f)$, with the condensate in the antisymmetric two-index
representation.
Confronting these with the composite-Higgs cosets: the minimal $\SO(5)\to\SO(4)$ scenario
\cite{Agashe:2004rs} has no simple fermionic realization \cite{Caracciolo:2012je};
$\SU(4)\to\SO(4)$ gives nine Goldstones in a $(3,3)$ with no Higgs candidate but underlies
minimal walking Technicolor \cite{Sannino:2004qp,Dietrich:2006cm,Foadi:2007ue,Frandsen:2009mi}
and its lattice studies \cite{Hietanen:2012sz,Hietanen:2013gva,Rantaharju:2017eej,Rantaharju:2019nmh}
(the two-Higgs-doublet variant of \cite{Mrazek:2011iu} requires an adjoint $15$, not a fermion
condensate); $\SO(6)\sim\SU(4)\to\Sp(4)\sim\SO(5)$, of interest for (ultra)minimal Technicolor
\cite{Appelquist:1999dq,Duan:2000dy,Ryttov:2008xe} and the composite Goldstone Higgs
\cite{Katz:2005au,Gripaios:2009pe,Galloway:2010bp,Barnard:2013zea,Ferretti:2013kya,Lewis:2011zb,Hietanen:2013fya},
gives five Goldstones as $(2,2)+(1,1)$; $\SU(5)\to\SO(5)$ yields fourteen Goldstones
\cite{Katz:2005au,Ferretti:2013kya}; $\SU(6)\to\Sp(6)$ gives two doublets and six singlets
\cite{Katz:2005au,Cai:2018tet}; and the complex case $\SU(4)\times\SU(4)\to\SU(4)$
\cite{Ma:2015gra} yields fifteen Goldstones---two Higgs doublets, one triplet and four
singlets---and also admits a pNGB dark-matter candidate \cite{Ma:2017vzm}.

Hawking radiation is an ideal astrophysical environment to observe high energy particle production and interactions, complementing collider searches of new particles.

%=====================================================================================

\subsection{MSSM scenarios}
It is known that the free parameter space to define a MSSM scenario is very large. The well-established approach to handle this, is to reduce the analysis to a smaller range of relevant parameters, such that one can bound step by step the allowed space by confronting both experimental results at Large Hadron Collider \cite{LHC_Evans:2008zzb, ATLAS:2022ihe, CMS:2019pov} and theoretical bounds. The last updated benchmark scenarios compatible with the Higgs sector of the SM  are reviewed in \cite{bagnaschi2019mssm, bahl2019mssm}, where one can find a list of six MSSM benchmarks constrained by LHC searches. In these scenarios, at least one of the supersymmetric Higgs bosons has properties compatible with the SM Higgs.

The most natural MSSM scenario is the $M_{h}^{125}$, as it predicts all the \SuSy\ particles to be heavy (masses above the TeV range) and the 125 GeV Higgs mass is associated to the lightest particle $h^0$ of the \SuSy\ Higgs sector. In addiction, the predicted masses of the third-generation squarks and the gluino are safely above the current bounds from direct searches at the LHC. The \SuSy\ input parameters for \texttt{SPheno} \cite{porod2012spheno} are:
\begin{equation} \begin{array}{ll} M_{Q_3} = M_{U_3} = M_{D_3} = 1.5~\text{TeV}, & M_{L_3} = M_{E_3} = 2~\text{TeV}, \\[6pt] \mu = 1~\text{TeV}, \quad M_1 = 1~\text{TeV}, & M_2 = 1~\text{TeV}, \quad M_3 = 2.5~\text{TeV}, \\[6pt] X_t = 2.8~\text{TeV}, & A_b = A_\tau = A_t \, . \end{array} 
\end{equation}
In this scenario there are two free parameters, such that theoretical and phenomenological searches can be presented in a two-dimensional plane $(\tan\beta, M_A)$. With reference to Fig. 1 in \cite{bagnaschi2019mssm}, it is clear that the only allowed region compatible with a Higgs mass of $125$~GeV is the horizontal line with $\tan\beta \simeq 20$ and $1200~\text{GeV} \leq M_A \leq 2000~\text{GeV}$ (which is a realistic upper bound).
As we would like the result to be as general as possible we scanned the parameter space in different points to check how much the results depended on the choice. 
Different runs with \texttt{SPheno v3.1} \cite{porod2012spheno} show that, for $\tan\beta = 20$ and different $M_A$ choices 
(1200 GeV, 1600 GeV and 2000 GeV), the resulting couplings, masses and branching ratios 
display only negligible differences among the three cases. 
The only sector that is more affected within this $M_A$ range is the Higgs sector, 
where one can observe a shift of up to $\sim 800$ GeV in the heavy Higgs masses 
(while preserving the same mass hierarchy), and the opening at $M_A=2000$ GeV 
of the decay channels of $A^0$ into neutralinos and charginos. 

Despite these features could mildly affect the primary emission probabilities during BH evaporation, two important remarks must be made: first, the BH temperature  
in the final evaporation stages is much higher than the scale of a few hundred GeV, 
so mass differences of this order can be neglected; second, the new decay channels into neutralinos 
and charginos open only at $M_A=2000$ GeV and have relatively small branching ratios compared to the fermionic decays (which, however, are the less  efficient channels for secondary photon production). In general, the heavy Higgs sector is not dominant in secondary photon production. The main source of photons comes from production of quarks and squarks which will hadronize in neutral pions $\pi^0$ with a $\simeq99\%$ BR into photon decays, so these differences are not expected to significantly affect 
the final photon spectra.  
For what concerns neutrinos, the comparison among the three choices of $M_A$ is more 
subtle. In fact, while the additional decay channels of $A^0$ into charginos and 
neutralinos remain relatively small, their contribution to the final neutrino production is not negligible. This implies that the results for neutrinos are more sensitive 
to the choice of the benchmark point. 
Operatively we set $M_A=2000$ to compare our results with previous works like~\cite{baker2023black}. 
While a direct check through \texttt{Pythia} 
would be useful to confirm the previous semi-quantitative considerations, in the following the $M_A=2000$ GeV scenario 
will be considered as representative of SuSy $M_{h}^{125}$ photon emission.

Another interesting benchmark studied in \cite{bagnaschi2019mssm} is the $M_{\tau}^{125}$ providing the same mass hierarchy in the Higgs sector, but predicting lighter charginos, neutralinos and third-generation fermions, which can influence the Higgs phenomenology, providing, for example, direct decays of the Higgs bosons into superparticles. The input parameters for this scenario are: 
\begin{equation}
\begin{array}{ll}
M_{Q_3} = M_{U_3} = M_{D_3} = 1.5~\text{TeV}, 
& M_{L_3} = M_{E_3} = 350~\text{GeV}, \\[6pt]
\mu = 1~\text{TeV}, \quad M_1 = 180~\text{GeV}, 
& M_2 = 300~\text{GeV}, \quad M_3 = 2.5~\text{TeV}, \\[6pt]
X_t = 2.8~\text{TeV}, \quad A_b = A_t, 
& A_\tau = 800~\text{GeV} \, .
\end{array}
\end{equation}
Further details on these two benchmark scenarios can be found in \cite{bagnaschi2019mssm}.

%=====================================================================================
\subsection{NMSSM scenario}
The NMSSM is a minimal
extension of the MSSM, obtained by
adding one gauge-singlet chiral superfield \(S\).
The main motivation for this extension is the so-called \(\mu\)-problem of the
MSSM: the supersymmetric Higgs mass parameter \(\mu\), appearing in the
superpotential term \(\mu H_u H_d\), is required phenomenologically to be of
the order of the EW scale, although its origin is not explained within
the MSSM.

In the NMSSM, the explicit \(\mu\)-term is replaced by a dynamical interaction
involving the singlet field. The scale-invariant superpotential can be written
as
\begin{equation}
W_{\rm NMSSM}
=
W_{\rm Yuk}
+
\lambda S H_u H_d
+
\frac{\kappa}{3} S^3 ,
\end{equation}
where \(W_{\rm Yuk}\) contains the usual Yukawa interactions of the SM fermions, while \(\lambda\) and \(\kappa\) are dimensionless couplings.
When the scalar component of the singlet field acquires a vacuum expectation
value, an effective \(\mu\)-parameter is generated dynamically,
\begin{equation}
\mu_{\rm eff} = \lambda \langle S \rangle .
\end{equation}
In this way, the EW size of the \(\mu\)-term is related to the
spontaneous breaking of the extended scalar sector rather than being inserted
by hand.

The additional singlet modifies the phenomenology of both the Higgs and
neutralino sectors. In the Higgs sector, the NMSSM contains two additional
neutral scalar degrees of freedom with respect to the MSSM. In the
CP-conserving case, the physical Higgs spectrum consists of three CP-even
neutral Higgs bosons, two CP-odd neutral Higgs bosons, and one charged Higgs
pair:
\begin{equation}
H_1, H_2, H_3, \qquad A_1, A_2, \qquad H^\pm .
\end{equation}
This enlarged Higgs sector allows for a richer phenomenology than in the MSSM,
including the possibility of light singlet-like scalar or pseudoscalar states.

The neutralino sector is also enlarged. Besides the four MSSM neutralinos, the
fermionic component of the singlet superfield introduces an additional
neutralino state, usually called the singlino. If the singlino has a large
singlet component, its couplings to Standard Model particles can be suppressed.
In scenarios where the singlino is the lightest supersymmetric particle, it can
play the role of a stable weakly interacting state, provided \(R\)-parity is
conserved. The NMSSM therefore provides a simple and well-motivated framework
in which the solution to the MSSM \(\mu\)-problem is accompanied by distinctive
Higgs and neutralino phenomenology.

For the NMSSM case, we can have an Higgs sector with a lighter higgs state. These scenarios are more complicated than the standard MSSM case, as the parameter space cannot be reduced to a two-dimensional plane. The LHCHXSWG3 provides different viable benchmarks for this scenario. We choose the BP7\_1 benchmark point, to have an Higgs sector with the  lightest Higgs about 96 GeV and most of the supersymmetric particles around 1 TeV. We determined the full spectrum with the specialized software \texttt{NMSSMTools v6.1.2} \cite{Ellwanger:2004xm, Ellwanger:2005dv}.

%==================================================================================

\subsection{A Composite Higgs benchmark}
 
To properly model the hadronisation due to the \CoDy\ interactions, we focus on a template model with properties as close as possible to QCD: the minimal global symmetry is  $\SU(4)\times\SU(4)\to\SU(4)$, as generated by a gauge symmetry with fermions in a complex representation. This complex-representation coset is especially rich~\cite{Ma:2015gra}, accommodating a pNGB spectrum consisting of two Higgs doublets, two custodial triplets and one singlet, and a built-in symmetry that makes one of them a good dark-matter candidate~\cite{Ma:2017vzm}. All of this while still allowing the custodial $\SO(4)$ needed for a realistic composite Higgs model.

In our implementation, the underlying dynamics consists of a
new confining gauge interaction, \(SU(3)_{\rm CH}\), with four Dirac fermions
\(\psi_i\) in the fundamental representation. The EW group is embedded in the global
flavour symmetry via the EW quantum numbers of the fermions, in such a way that the custodial symmetry is preserved:
two fermions, say $\psi_1$ and $\psi_2$, form an \(SU(2)_L\) doublet, while the other two, $\psi_3$ and $\psi_4$, form an \(SU(2)_R\) doublet (where hypercharge is identified with the diagonal
generator of \(SU(2)_R\)).
In the infrared regime, the strong dynamics generates a fermion condensate that breaks the
global symmetry as
\begin{equation}
SU(4)_1 \times SU(4)_2 \longrightarrow SU(4)_D \,,
\end{equation}
without breaking the custodial $SU(2)_L\times SU(2)_R$ subgroup. The corresponding coset contains fifteen pNGBs,
transforming under the custodial group
\(SU(2)_L\times SU(2)_R\) as
\begin{equation}
\mathbf{15}
=
\underbrace{(2,2)}_{H_1}
+
\underbrace{(2,2)}_{H_2}
+
\underbrace{(3,1)}_{\Delta}
+
\underbrace{(1,3)}_{N}
+
\underbrace{(1,1)}_{s}\,,
\end{equation}
where the two bidoublets \(H_1\) and \(H_2\) play the role of two Higgs doublets,
\(\Delta\) and \(N\) are triplets, and \(s\) is a singlet.
%Therefore, the low-energy scalar spectrum contains two electroweak Higgs doublets, two custodial triplets and one singlet. 
This realizes a composite two-Higgs-doublet
model, where the Higgs fields arise as pNGBs of the strong sector rather than as
elementary scalars. Generating a vacuum expectation value for the two doublets creates the vacuum misalignment that finally breaks the EW symmetry at a scale $v = f \sin \theta$, where $f$ is the decay constants of the pNGBs. It was realized that the $\tan \beta$ parameter, which parametrizes how $v$ is distributed between the two doublets (c.f. MSSM), is unphysical due to the extended global symmetry of the coset~\cite{Ma:2015gra}. Besides the Higgs pNGB states, the theory generically predicts additional
composite resonances with masses around the confinement scale, typically in the multi-TeV
range, potentially providing characteristic signatures in the Hawking radiation spectra. In fact, as the BH evaporates, it will explore two different regimes: the confined IR regime of the composite sector, when $T_{\mathrm{BH}} \lesssim \Lambda_{\mathrm{CH}}$ and only the new bound states can be emitted; the UV asymptotically free regime, when $T_{\mathrm{BH}}\gtrsim  \Lambda_{\mathrm{CH}}$. In the latter, the \CoDy\ elementary fermions and gluons can be resolved and they will undergo hadronization and showering into the new composite sector. As detailed in Section~\ref{sec:methodology}, the details of these non-perturbative processes are modelled via the Hidden Valley module of \texttt{Pythia}.

In our analysis, for simplicity, we only study the production of pNGBs via hadronization~\cite{Cacciapaglia_2025}. The decays of the pNGBs, other than the SM-like Higgs, crucially depend on the operators generating the top quark mass.
In the effective Lagrangian the Yukawa couplings can be written in the form~\cite{Ma:2015gra}:
\begin{equation}
\begin{aligned}
\mathcal{L}_{\rm Yuk}
=&-\left[
Y_t v_{\rm SM}
+Y_t\cos\theta\, h_1
+iY_D h_2
+Y_D\cos\theta\, A_0
+i\frac{Y_T}{\sqrt{2}}\sin\theta\,(N_0+\Delta_0)
\right]
\left(\bar{\chi}_{t_L}\bar{\eta}_{t_R}\right)
\\
&-\left[
-i\sqrt{2}Y_D\cos\theta\, H^-
+iY_T\sin\theta\,(N^-+\Delta^-)
\right]
\left(\bar{\chi}_{b_L}\bar{\eta}_{t_R}\right)
+\text{h.c.}
\end{aligned}
\label{eq:yukawa_expansion}
\end{equation}
which contains three independent couplings: $Y_t$, $Y_T$ and $Y_D$,
We interpret them in the following way:
\(Y_t\) corresponds to the effective top Yukawa coupling, with
\(m_t=Y_t v_{\rm SM}\), \(Y_D\) is the coupling of the second doublet to the top,
and \(Y_T\) is the coupling of the two triplets. For models with $Y_D = Y_T = 0$, the second doublet and triplets are odd under a built-in parity, that renders the lightest state a good Dark Matter candidate~\cite{Ma:2017vzm}. In the following, instead, we will consider a benchmark with non-zero $Y_D$ and $Y_T$
, so that the couplings in Eq.~\eqref{eq:yukawa_expansion} derive the decays of all pNGBs, except for the singlet $s$. For the latter, one should also consider a Wess-Zumino-Witten term in the effective Lagrangian to account for triangle loop anomaly. Such term reads:
\begin{equation}
\mathcal{L}_{WZW}
\sim
k\, \cos \theta\, \frac{s}{f}
\left(
g^2 W_{\mu\nu}\widetilde{W}^{\mu\nu}
-
g'^2 B_{\mu\nu}\widetilde{B}^{\mu\nu}
\right)\,.
\end{equation}
The Yukawa and WZW terms can be exploited to determine the possible decay channels of the pNGBs and the corresponding branching ratios. It is clear from Eq. (\ref{eq:yukawa_expansion}) that the doublets and triplets have unique decay channels into top and bottom quarks, while the singlet $s$ can undergo different decays, as listed in Table~\ref{tab:benchmark_particle_spectrum_decays}.

\begin{table}[htbp]
\centering
\small
\renewcommand{\arraystretch}{1.35}
\begin{tabular}{c c c c c c c}
\toprule
\textbf{Particle} &
\textbf{Mass [GeV]} &
\textbf{Q} &
\textbf{Spin} &
\textbf{d.o.f.} &
\textbf{Decay channels} &
\textbf{BR} \\
\midrule

\(W^\pm\)
& \(80.4\)
& \(\begin{array}{c}
+1 \\
-1
\end{array}\)
& \(1\)
& \(6\)
& SM decays
& SM-like
\\
\midrule

\(Z\)
& \(91.188\)
& \(0\)
& \(1\)
& \(3\)
& SM decays
& SM-like
\\
\midrule

\(h_1\)
& \(125\)
& \(0\)
& \(0\)
& \(1\)
& SM decays
& SM-like
\\
\midrule

\(h_2\)
& \(800\)
& \(0\)
& \(0\)
& \(1\)
& \(h_2 \to t\bar{t}\)
& \(100\%\)
\\
\midrule

\(A_0\)
& \(800\)
& \(0\)
& \(0\)
& \(1\)
& \(A_0 \to t\bar{t}\)
& \(100\%\)
\\
\midrule

\(H^\pm\)
& \(800\)
& \(\begin{array}{c}
+1 \\
-1
\end{array}\)
& \(0\)
& \(2\)
& \(\left\{
\begin{array}{l}
H^+ \to t\bar{b} \\
H^- \to \bar{t}b
\end{array}
\right.\)
& \(\begin{array}{c}
100\% \\
100\%
\end{array}\)
\\
\midrule

\(N_0\)
& \(800\)
& \(0\)
& \(0\)
& \(1\)
& \(N_0 \to t\bar{t}\)
& \(100\%\)
\\
\midrule

\(N^\pm\)
& \(800\)
& \(\begin{array}{c}
+1 \\
-1
\end{array}\)
& \(0\)
& \(2\)
& \(\left\{
\begin{array}{l}
N^+ \to t\bar{b} \\
N^- \to \bar{t}b
\end{array}
\right.\)
& \(\begin{array}{c}
100\% \\
100\%
\end{array}\)
\\
\midrule

\(\Delta_0\)
& \(800\)
& \(0\)
& \(0\)
& \(1\)
& \(\Delta_0 \to t\bar{t}\)
& \(100\%\)
\\
\midrule

\(\Delta^\pm\)
& \(800\)
& \(\begin{array}{c}
+1 \\
-1
\end{array}\)
& \(0\)
& \(2\)
& \(\left\{
\begin{array}{l}
\Delta^+ \to t\bar{b} \\
\Delta^- \to \bar{t}b
\end{array}
\right.\)
& \(\begin{array}{c}
100\% \\
100\%
\end{array}\)
\\
\midrule

\(s\)
& \(800\)
& \(0\)
& \(0\)
& \(1\)
& \(\left\{
\begin{array}{l}
s \to W^+W^- \\
s \to Z\gamma \\
s \to ZZ
\end{array}
\right.\)
& \(\begin{array}{c}
60.8\% \\
35.5\% \\
3.7\%
\end{array}\)
\\

\bottomrule
\end{tabular}
\caption{Particle spectrum benchmark used in this work. The masses of the
pNGBs are fixed to \(800~\mathrm{GeV}\), while the gauge bosons properties are same as in the SM. The decay channels of the composite pNGBs include tree-level processes and 1-loop triangle anomaly.}
\label{tab:benchmark_particle_spectrum_decays}
\end{table}

As can be seen from the effective Lagrangian in Eq.~\eqref{eq:yukawa_expansion}, most of the pNGBs decay predominantly into top and bottom quarks. The subsequent decay of the latter into lighter quark states, followed by hadronization, leads to the production of light mesons, in particular, pions. Since pion decays provide an important secondary source for both photons and neutrinos, the presence of the composite pNGB sector is expected to affect their final spectra. Therefore, even when restricting the analysis to the confined regime of the model, one should already expect a non-negligible impact on the final particle spectra.

For the numerical benchmark, we choose a value for the decay constant $f \sim 1$~TeV in such a way to have the new heavy state masses around the TeV scale. In particular, we fix the masses of all pNGBs (except for the Higgs and longitudinal $W$ and $Z$ components) to be $800$~GeV, which should be above the current exclusion at the LHC for direct production~\cite{Cacciapaglia:2022bax}. We treat the fundamental fermions, together with the
gauge bosons mediating the new strong interaction, as additional degrees of freedom
above the confinement scale \(\Lambda_{CH} \sim 4 \pi f\), which we fix to be $14$~TeV. Their primary Hawking emission becomes
efficient when the BH temperature is comparable to their mass. Since the
low-energy confined description does not uniquely determine the effective masses of the
fundamental constituents, we adopt a benchmark value for the masses of the new sector
fundamental particles. In the following, we denote them by \(q_{\rm CH}\) and
\(g_{\rm CH}\), in analogy with quarks and gluons, and set
\begin{equation}
m_{q_{\rm CH}} = m_{g_{\rm CH}} = 2~\mathrm{TeV}
\end{equation}
and degrees of freedom (d.o.f.)
\begin{equation}
   \mathrm{d.o.f}(\mathrm{q_{CH}})=4 \times 3 \times 4 = 48 \quad, \quad  \mathrm{d.o.f}(\mathrm{g_{CH}})=2\times 8=16
\end{equation}
in the numerical simulation. The additional factor 4 in d.o.f($\mathrm{q_{CH}}$) accounts for the fact that in this model the four quarks are identical particles.
The full set of composite masses and particle properties used in the model is summarized in Table~\ref{tab:benchmark_particle_spectrum_decays}.

%=====================================================================================
\subsection{Results}
%To obtain the results, we used \texttt{BlackHawk v3.0}, which is a tool that can compute the primary and secondary spectra of stable particles generated by Hawking radiation, and in some cases can follow their evolution over time. The latest version of the code only includes the Standard Model particles as elementary particle content, so we needed to extend its capability to the different models we studied. In addiction, \texttt{BlackHawk v3.0} contains the implementation of different black hole metrics, including regularized ones, and its extension to SuSy and CoDy emission allows the exploration of different scenarios, where the modifications of the metric and the alternative particle model can intertwine. 

%To identify a reference threshold mass below which BSM degrees of freedom become relevant, we checked that significant deviations from the SM start to  appear when $\mathrm{M_{BH}} \sim 10^{10}\,\mathrm{g}$, corresponding to a Hawking temperature of about $1\,\mathrm{TeV}$, i.e., around the thermal threshold for BSM particles in the chosen benchmarks.In the composite Higgs scenario, due to the new confining regime, all the primary bound states are emitted up to a temperature $\mathrm{T_H \lesssim \Lambda_{\rm CH}}$, which then decays back to SM particles following the decay channels in Table \ref{tab:benchmark_particle_spectrum_decays}; in the opposite regime, $\mathrm{T_H\gtrsim \Lambda_{CH}}$, elementary CH states are promptly emitted by the black hole and undergo hadronization to form the composite heavy states.

\begin{figure}[h!]
    \centering

    \begin{subfigure}[t]{0.49\linewidth}
        \centering
        \includegraphics[width=\linewidth]
        {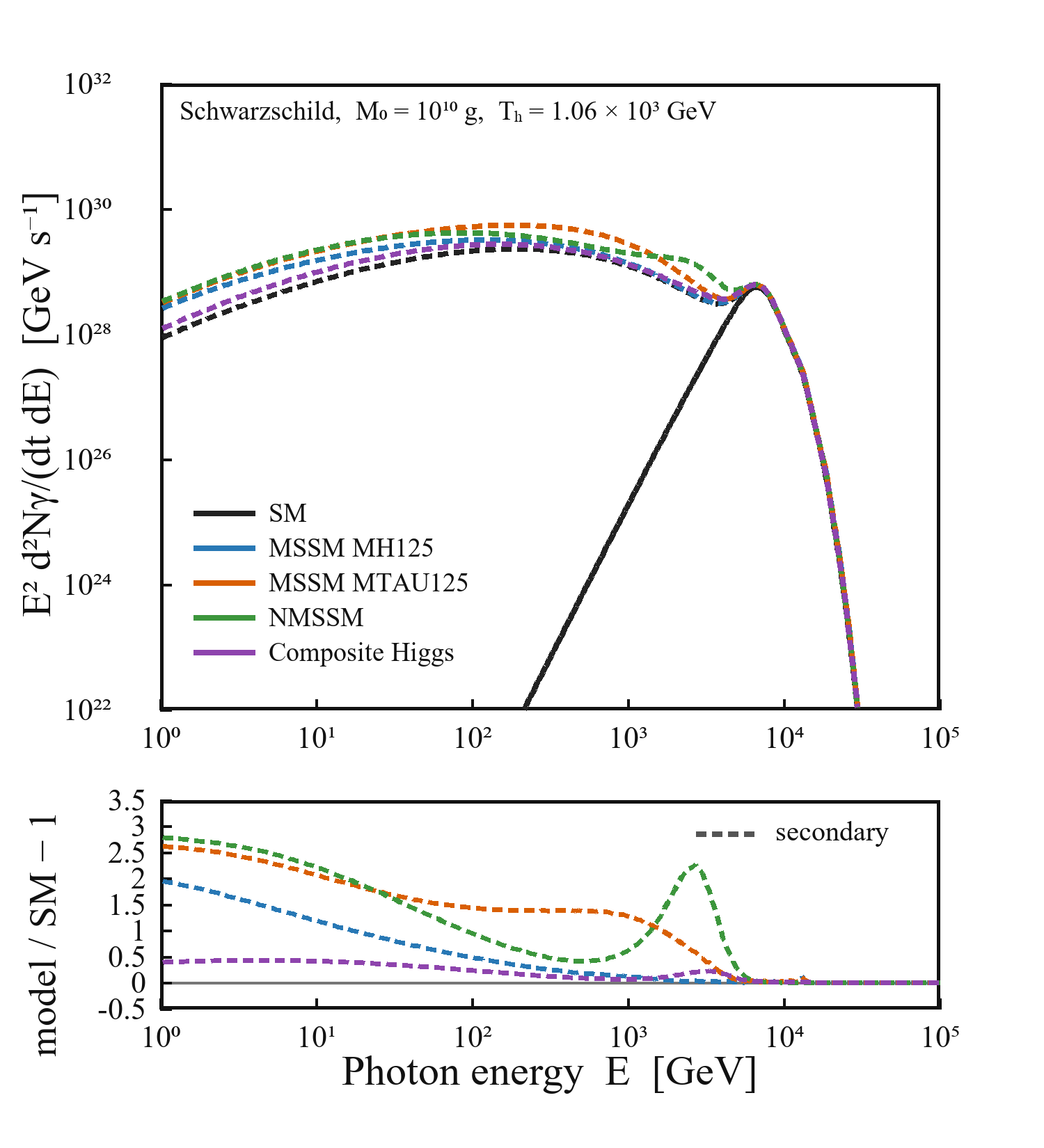}
        \caption{Schwarzschild, t=0}
        \label{fig:photon_SuSy_a}
    \end{subfigure}
    \hfill
    \begin{subfigure}[t]{0.49\linewidth}
        \centering
        \includegraphics[width=\linewidth]
        {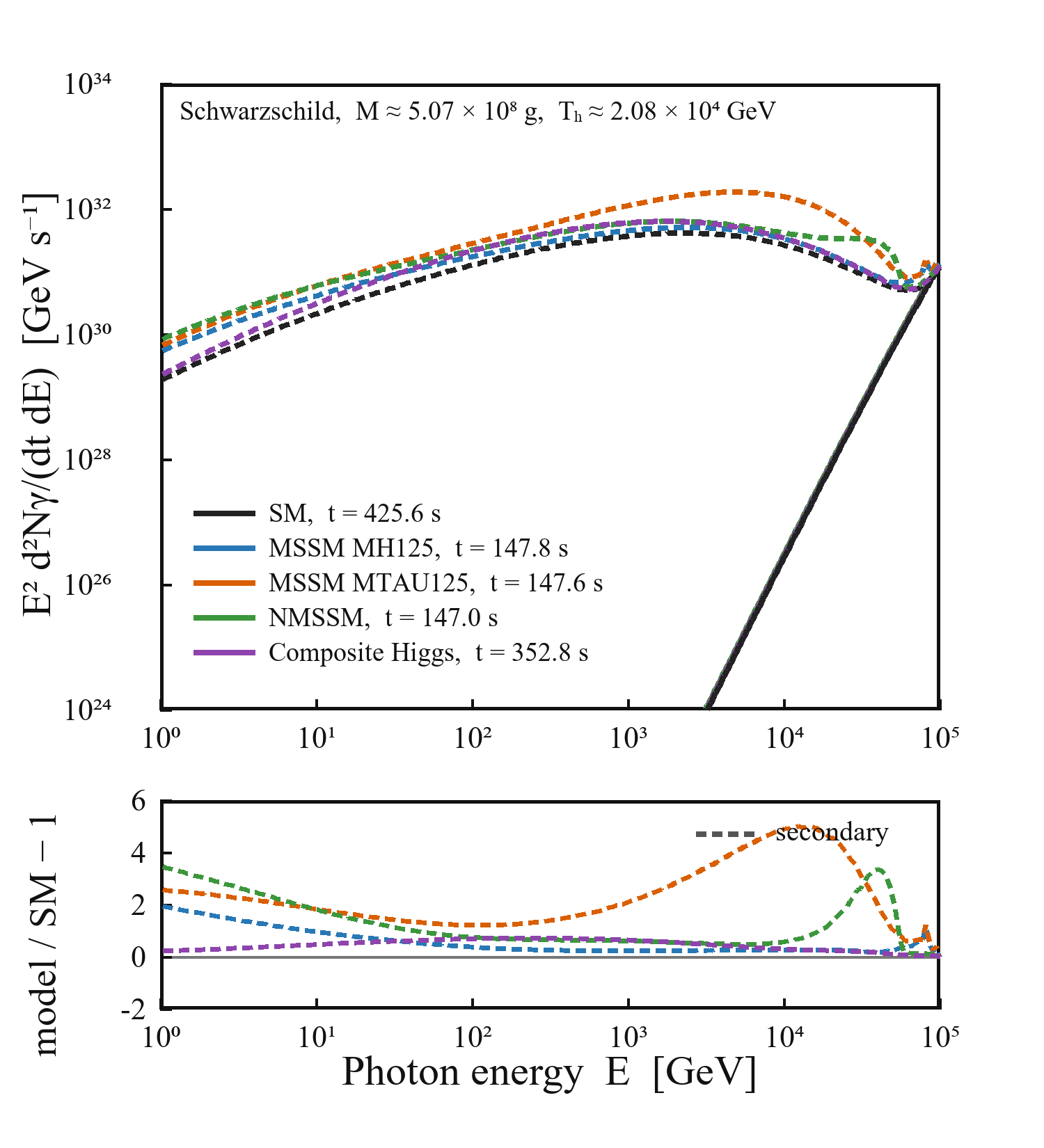}
        \caption{Schwarzschild, near-final}
        \label{fig:photon_SuSy_b}
    \end{subfigure}

    \caption{Comparison between the final photon produced in the SM versus the different scenairos considered. The lower panels show the ratio defined in Eq.(\ref{eq:ratio}).}
    \label{fig:photon_SuSy}
\end{figure}

As detailed in Section~\ref{sec:methodology}, the photon and neutrino spectra were generated thanks to the public code \texttt{BlackHawk v3.0}~\cite{Arbey:2026koc}, where we added in a private version the BSM benchmarks detailed above. 
The photons spectra are presented in Figure~\ref{fig:photon_SuSy} for all benchmark, while we refer to the Appendix~\ref{sec:appendix} for neutrino results. The spectra are computed instantaneously for two BH masses: $M_\text{BH} = 10^{10}$~g in the left panel (corresponding to $T_H \sim 1$~TeV), and $M_\text{BH} = 5\times 10^{8}$~g in the left panel (corresponding to $T_H \sim 20$~TeV).
%In Figure(\ref{fig:photon_CH}) we separate the effect of the composite model to better show the features of the specturm. To isolate the dependence on the particle model, the reference value $M_{BH}=10^{10}\,\mathrm{g}$ for the initial BH mass is adopted throughout, keeping in mind that lighter masses correspond to more intense and energetic emissions. 
Each spectrum is presented within two sub-panels: the upper ones contains the primary photons emitted by a Schwarzschild BH (black solid line), as well as the secondary photons produced in different particle physics models (dotted colored lines); the lower ones show the ratio 
\begin{equation}
R_{T_{\rm H}}(E)
=
\frac{
\left(d^{2}N/dt\,dE\right)_{\rm model}
}{
\left(d^{2}N/dt\,dE\right)_{\rm SM}
}
-1.
\label{eq:ratio}
\end{equation}
which better describes quantitatively the production differences of the photons in the BSM scenarios with respect to the SM. 

In fact, some interesting features emerge from the analysis of such plots. Let's first focus on Figure~\ref{fig:photon_SuSy}a, showing the secondary photons produced in the three different SuSy benchmarks and the CoDy one from a Schwarzschild black hole of initial mass $\mathrm{M_0}=10^{10}\mathrm{g}$, corresponding to an initial temperature around $\mathrm{T_H}\sim 1$ TeV. %An analogue plot for the CH scenario is presented in Figure (\ref{fig:photon_CH}a). 
In the ratio panel, we can distinguish some special features of the different benchmarks. The $\mathrm{M_h^{125}}$ plot, shows a continuous overproduction of photons, which increases at lower energies, in contrast with the other benchmarks ($\mathrm{M_{\tau}^{125}}$, the NMSSM and the CH), which manifest a bump in some energy windows. Additionally, by looking at  Figure~\ref{fig:photon_SuSy}b, we can see that, at higher temperatures, two different aspects can be distinguished: the plot remains quite unchanged apart from a translation towards higher energies (as in the $\mathrm{M_h^{125}}$ and NMSSM cases); or the morphology of the plot changes more dramatically, showing a big increase in the ratio amplitude, as in the $M_{\tau}^{125}$ case.

 To better understand the origin of such features it is important to take into account the mass spectra for the BSM particle in the different benchmarks, and their allowed decay channels. In the SM, the main sources for photon production are quarks and gluons, through hadronization and showering, which contribute to the low-energy tail of the spectra. At higher energies, the EW sector contributes to photon production through multiple decay channels. If a parent primary particle mass is around the EW energy scale ($100-200$~GeV), it is already efficiently produced when $\mathrm{T_H \sim 1\,\mathrm{TeV}}$, and this effect increases with the temperature. This is confirmed by an overall increment of the flux when $\mathrm{T_H \sim 20\,\mathrm{TeV}}$, and such feature is common to all particle models. The BSM states can be additional sources of secondary photons both through direct decay and through multiple decay channels. In the first case, the presence of a direct decay manifests as a narrow bump in the ratio plot, corresponding to the energy of the boosted photon.
  In addition to that, if the BSM states that contribute to the photon signal are around the EW masses, their increased efficiency will scale the same as the SM during the evaporation, therefore the relative number of additional photons in the BSM case remains approximately constant with the Hawking temperature. An example for this mechanism is the NMSSM case: here, we can see a narrow bump in the ratio plot, likely generated by the direct radiative decay
\begin{equation}
    \widetilde{\chi}_2^0
    \rightarrow
    \widetilde{\chi}_1^0 \gamma,
    \qquad
    \mathrm{BR} \simeq 62.9\% .
\end{equation}
 In this benchmark, $m_{\widetilde{\chi}_2^0} \simeq 128~\mathrm{GeV}$ and $m_{\widetilde{\chi}_1^0} \simeq 82~\mathrm{GeV}$, so the decay is kinematically allowed. In addition, $m_{\widetilde{\chi}_2^0} $ is of the same order as the electroweak-scale masses of the relevant SM parent particles. Therefore, both the emission of $\widetilde{\chi}_2^0$ and that of the SM parents are already efficient at $T_{\mathrm{H}} \simeq 1~\mathrm{TeV}$ and they scale approximately in the same way when the temperature increases. This explains why the bump shifts towards higher photon energies, while its height in the ratio doesn't increase drastically.

\begin{figure}[h!]
    \centering

    \begin{subfigure}[t]{0.49\linewidth}
        \centering
        \includegraphics[width=\linewidth]
        {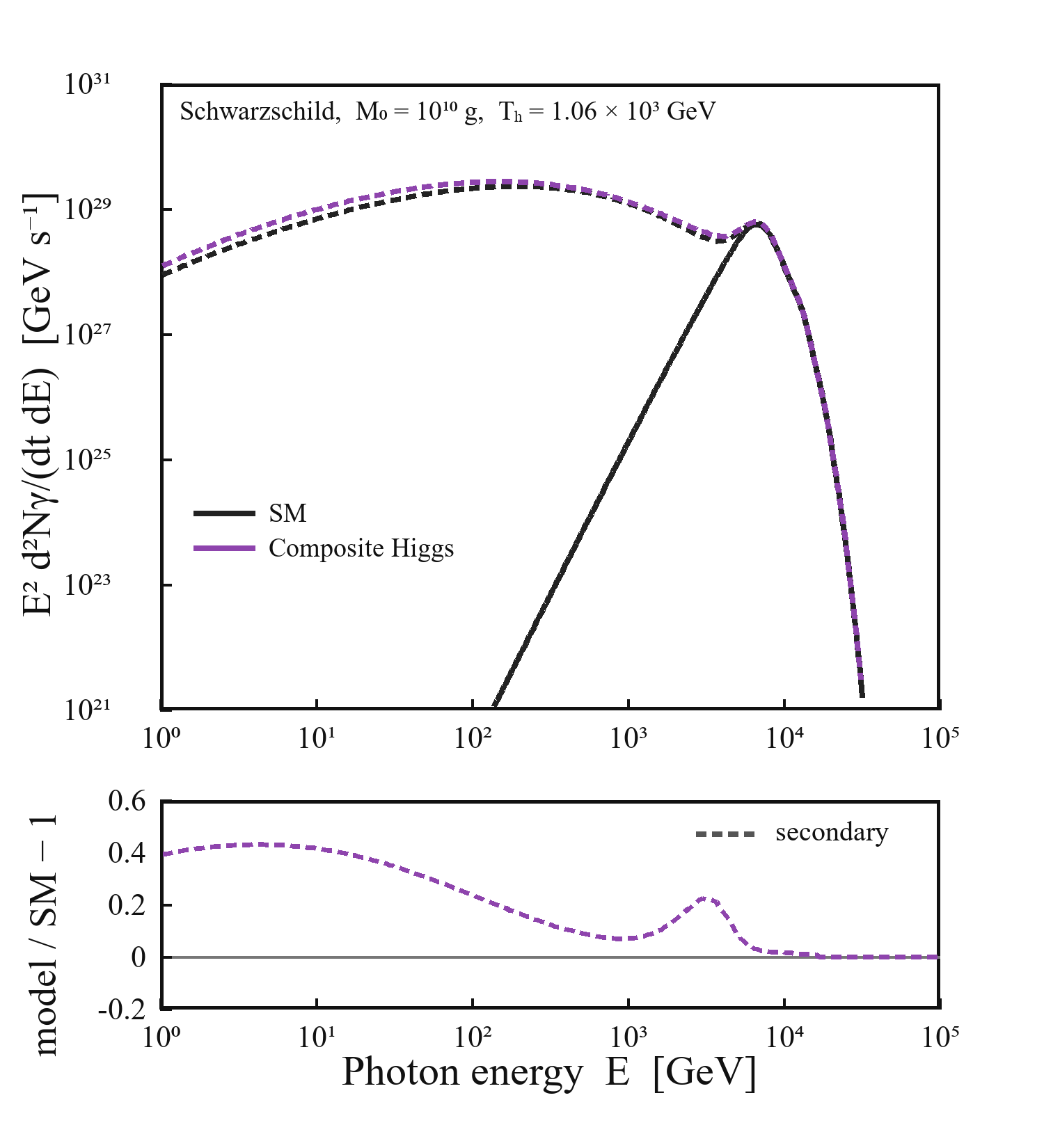}
        \caption{Schwarzschild, t=0}
        \label{fig:photon_CH_a}
    \end{subfigure}
    \hfill
    \begin{subfigure}[t]{0.50\linewidth}
        \centering
        \includegraphics[width=\linewidth]
        {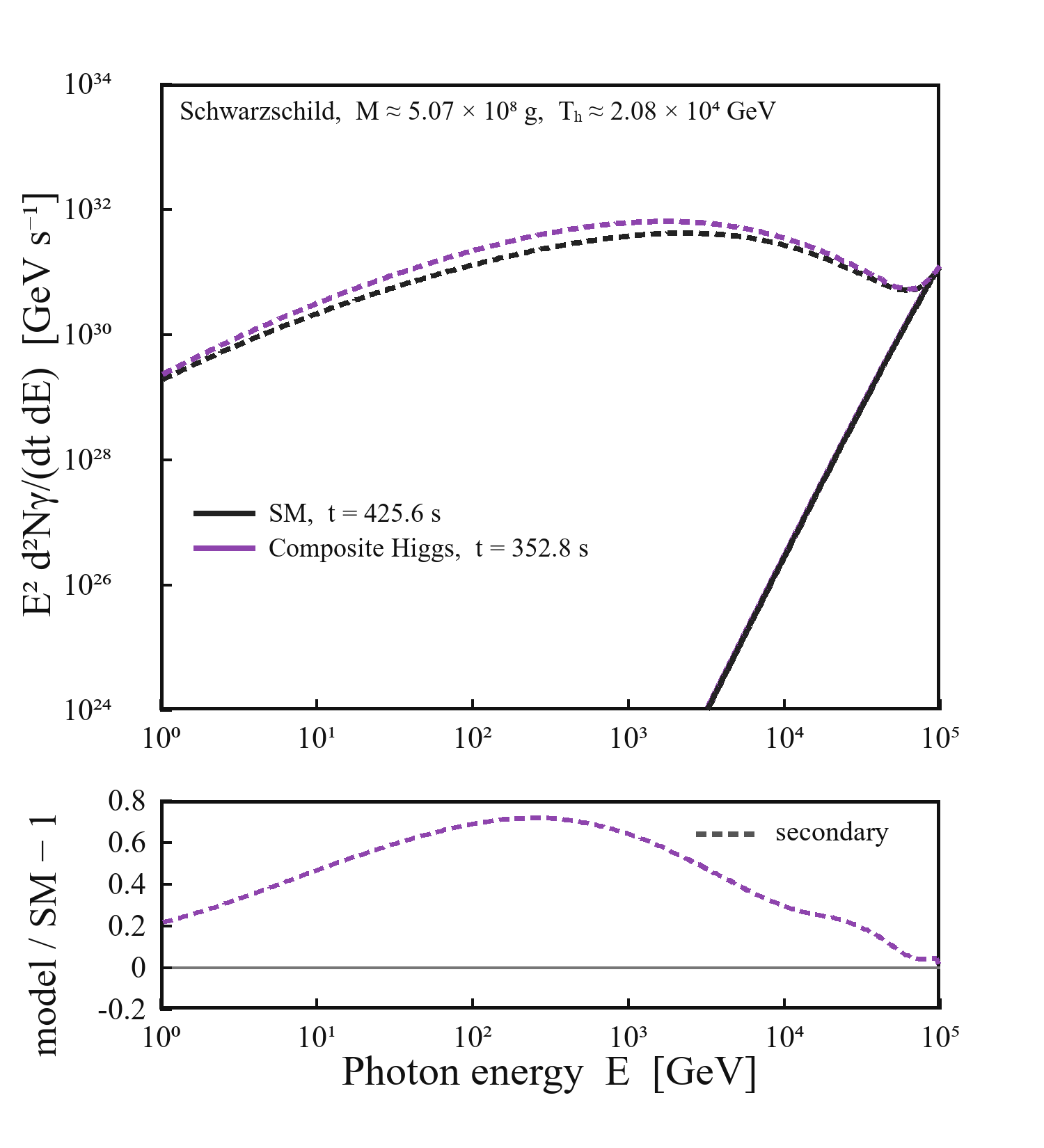}
        \caption{Schwarzschild, near-final}
        \label{fig:photon_CH_b}
    \end{subfigure}

    \caption{Isolated comparison between the final photon produced in the SM versus the Composite Higgs scenario. The BH in the near-final stages with $T_H \simeq 20 \rm TeV$is in the confined regime, where it remains for about 0.13s after the threshold.}
    \label{fig:photon_CH}
\end{figure}

A similar narrow feature appears in the CH benchmark, which we plot in isolation in Figure~\ref{fig:photon_CH} for clarity. The bump is due to the direct decay
\begin{equation}
    s \rightarrow Z\gamma , \qquad
    \mathrm{BR} \simeq 35.5\%
\end{equation}
which is generated with a sizeable branching ratio, C.f. Table~\ref{tab:benchmark_particle_spectrum_decays}.
However, in this case the parent particle has a larger mass as compared to the NMSSM case, $m_s \simeq 0.8~\mathrm{TeV}$. Consequently,
\begin{equation}
    \frac{m_s}{T_{\mathrm{H}}} \simeq 0.75
    \quad \text{at} \quad
    T_{\mathrm{H}} = 1.06~\mathrm{TeV},
    \qquad
    \frac{m_s}{T_{\mathrm{H}}} \simeq 0.04
    \quad \text{at} \quad
    T_{\mathrm{H}} = 20~\mathrm{TeV}
    \end{equation}
The emission of the composite parent, therefore, becomes much more efficient as the temperature increases, while the relevant SM particles were already efficiently emitted at the lower temperature. Its contribution consequently grows more rapidly than the SM photon spectrum, increasing the height of the bump, as shown in the bottom panel of  Figure~\ref{fig:photon_CH}b as compared to Figure~\ref{fig:photon_CH}a.

A similar temperature-dependent enhancement occurs in the $M_{\tau}^{125}$ benchmark. In this case, the particles that become relatively more important at $T_{\mathrm{H}} = 20~\mathrm{TeV}$ have masses of approximately $1$--$2.5~\mathrm{TeV}$. Their Hawking emission therefore increases more rapidly than that of the electroweak-scale SM parents. However, $M_{\tau}^{125}$ does not contain a relevant direct photon channel. The heavy states instead decay through several combinations of $W$, $Z$, $h$, top quarks, and jets. The photons are consequently produced through many different decay chains and showers, resulting in a broad enhancement rather than a narrow bump. An important channel in the lighter sector is
\begin{equation}
    \widetilde{\chi}_2^0
    \rightarrow
    \widetilde{\chi}_1^0 h,
    \qquad
    \mathrm{BR} \simeq 85.5\% .
\end{equation}
The resulting photons are mainly produced through subsequent Higgs decays and showers, rather than directly through $h \rightarrow \gamma\gamma$.
Finally, no analogous bump appears in the $M_{h}^{125}$ benchmark. The decay
\begin{equation}
    \widetilde{\chi}_2^0
    \rightarrow
    \widetilde{\chi}_1^0 h
\end{equation}
is kinematically forbidden because
\begin{equation}
    m_{\widetilde{\chi}_2^0}
    -
    m_{\widetilde{\chi}_1^0}
    \simeq
    58.6~\mathrm{GeV}
    <
    m_h .
\end{equation}
The radiative decay
\begin{equation}
    \widetilde{\chi}_2^0
    \rightarrow
    \widetilde{\chi}_1^0 \gamma
\end{equation}
is kinematically allowed, but its branching ratio is only about
$\mathrm{BR} \simeq 1.1 \times 10^{-4}$ and is therefore negligible. Photon production is consequently dominated by multibody decays, final-state radiation, and particle showers, which generate a smooth continuum instead of a localized spectral feature.

Overall, the NMSSM bump is associated with a single dominant radiative decay 
whose relative contribution remains stable with temperature. The $\mathrm{M_{\tau}^{125}}$ instead 
develops a temperature-dependent broad enhancement due to the increasing 
importance of heavy primary states and multistep cascades. The $\mathrm{M_{h}^{125}}$ shows no 
localized feature because its compressed electroweak spectrum suppresses or 
kinematically forbids the relevant photon- and Higgs-producing channels.
The neutrino spectra can be interpreted in the same way, using analogous considerations based on the decay tables and mass spectra of the parent particles. The secondary neutrino spectra for the different models are shown in Appendix~\ref{sec:appendix}.
%=====================================================================================

%======================================================================================
\section{Evaporation from regularized black holes} \label{sec:RBHs}
The Hawking evaporation process can be strongly affected by the nature of the BH metric through the greybody factors $\Gamma(E,M)$. %A special class of BHs, called regular black holes, has gained a lot of interest recently as they are black hole solution with no singularity. In a complete theory of gravity, we expect BHs metric to be modified in such a way to be geodetically complete. In other words we expect BH to not present any spacetime singularity and to be describable by what we call a regular BH metric. 
Here, we want to focus on how the regularized metrics of RBHs can affect the standard Hawking evaporation and if there's some regime in which effects of BSM physics and ``quantum gravity'' can overlap. 
Hawking evaporation in some RBH metrics was investigated in \cite{Arbey:2026koc}. 
We decided to explore three different regularized models: a loop quantum gravity inspired model for polymerized BHs and the well known Hayward \cite{Hayward:2005gi} and Simpson-Visser \cite{Simpson:2018tsi} regularized metrics. These are chosen to be representative of the two possible classes of RBHs: models with a regular core surrounded by a double horizon structure (Hayward being the prototypical example, alongside the polymerized/LQG-inspired metric discussed below), and models in which the classical singularity is replaced by a wormhole throat hidden behind the horizon (the Simpson-Visser ``black-bounce'' metric being the reference case).

The metric of polymerized BHs is parameterized in terms of a small parameter $\varepsilon$.
Instead, the other two metrics are parametrized by a common parameter L~\cite{Arbey:2026koc}, defined as:
\begin{equation}
    L=\frac{\ell}{\mathrm{\ell_{max}(M)}} \qquad L \in [0,1)
    \label{reg_parameter}
\end{equation}
where $\ell$ fixes the scale for new gravitational physics, and $\ell_\mathrm{max}(M)$ is the maximal value of $\ell$ beyond which these metrics for a mass $M$ do not feature an horizon: hence, for $L>1$ (i.e. $\ell > \ell_\mathrm{max}(M)$) they simply describe a horizon-less compact object. The value of $\ell_\mathrm{max}(M)$ depends on the specific regularized model and on the BH mass. For $L=1$ (i.e. $\ell= \ell_{\text{max}}(M)$), the effective Hawking temperature is exactly 0. Indeed, the critical $L=1$ value corresponds to an extremal BH that stops evaporating and becomes a remnant (this also represents one of the viable solutions for the information loss paradox). %From Eq.~\eqref{reg_parameter}, it is clear that once we choose a regularized metric and a BH mass for instantaneous emission, $\ell_\mathrm{max}$ is fixed and the scale $\ell$ becomes upper constrained. This means that, if a BH starts emitting new particles efficiently when in reaches a certain mass M, the $\ell$ value must be tuned accordingly if one wants to see both effects on the final evaporation.

Before exploring the results, we will give a brief review of these models.

\subsection{Polymerized BHs}
The general form of spherically symmetric static metrics in four-dimensional Boyer-Lindquist coordinates is
\begin{equation}
 ds^2 = -G(r)dt^2 + \frac{1}{F(r)}dr^2 + H(r)d\Omega^2
\end{equation}
where $d\Omega^2 = d\theta^2 + \sin^2\theta d\phi^2$ is the metric on the unit 2-sphere.
 In Loop Quantum Gravity  (LQG) \cite{Rovelli_1998}, quantum gravity effects can be encoded in effective semiclassical corrections to a classical BH metric (often a static Schwarzschild BH). The metrics derived in this way describe the so-called \textit{polymerized black holes}. There are many proposals for deriving such BH metrics \cite{Zhang:2023yps}, depending on the quantization scheme of the loop Schwarzschild BH. The simplest case is the $\mu_0$-scheme, in which the metric coefficients have the form
\begin{equation}
F(r) = \frac{(r - r_{+})(r - r_{-})r^{4}}{(r + r_{\ast})^{2}(r^{4} + a_{0}^{2})}, \quad 
G(r) = \frac{(r - r_{+})(r - r_{-})(r + r_{\ast})^{2}}{r^{4} + a_{0}^{2}}, \quad 
H(r) = r^{2} + \frac{a_{0}^{2}}{r^{2}},
\end{equation}
where 
\begin{equation}
r_{\rm H} \equiv r_{+} \equiv 2m(\varepsilon), \qquad r_{\rm C} \equiv r_{-} \equiv 2m(\varepsilon)P(\varepsilon)^{2},
\end{equation}
are the horizon and Cauchy radii. The two free parameters defining the metric are the LQG area gap $a_0$ (that we put to 0 in our simulations)  and $\varepsilon$, the latter representing the effective polymerization parameter and describing the scale of quantum corrections. This metric is asymptotically flat and recovers the Schwarzschild case in the limits $\varepsilon \to 0$ and $a_{0} \to 0$.

Polymerized BHs belong to a class of metric solutions that resolve the central singularity problem, as the metric coefficients $F(r)$, $G(r)$ and $H(r)$ do not diverge at $r \to 0$. 
The important quantities for the BH evaporation can be written in terms of a polymerization function $P(\epsilon)$
\begin{equation}
P(\varepsilon) = \frac{\sqrt{1 + \varepsilon^{2}} - 1}{\sqrt{1 + \varepsilon^{2}} + 1}.
\end{equation}
The parameter $m$ is related to the ADM (Arnowitt-Deser-Misner) mass $M$ by:
\begin{equation}
M = m(\varepsilon)(1 + P(\varepsilon))^{2}.
\end{equation}
and the temperature is given by:
\begin{equation}
T_{LQG} = \frac{r_+^2(r_+-r_-)}{4\pi(r_+^4+a_0^2)}.
\end{equation}
An important feature to highlight, is that the polymerization factor is a constant inherited from the quantum nature of gravity, and it should be thought as an ``effective'' parameter. %In presence of a complete theory, its content would be only linked to the quantum nature of gravity. 
Therefore,  $\varepsilon$ does not evolve during the whole BH lifetime and its presence impact the emission properties during all the evaporation stages of the BH. % (i.e., the decrease of the BH temperature compared to the Schwarzschild case, and smaller emission rates) characterize all evaporation stages.

\subsection{Hayward regularized solution}

The Hayward metric \cite{Hayward:2005gi} is a standard example of a static, spherically symmetric RBH. Its interior smoothly matches onto a de Sitter core, so that curvature invariants remain finite everywhere including $r=0$.
In Schwarzschild-like coordinates,
\begin{equation}
    ds^2 = -f(r)dt^2 + \frac{dr^2}{f(r)} + r^2 d\Omega^2,
\end{equation}
with
\begin{equation}
    f(r) = 1 - \frac{2Mr^2}{r^3 + 2\ell^2M},
\end{equation}
where $M$ is the ADM mass and $\ell$ is the regularization length scale controlling the curvature at the origin. In the limit $r\to0$, $f(r) \to 1 - r^2/\ell^2$, i.e. we have a de Sitter core with effective cosmological constant $\Lambda \sim 3/\ell^2$; in the limit $\ell\to0$ the Schwarzschild solution is recovered, while for $r\gg \ell$ the metric is asymptotically flat.

Depending on the ratio between $\ell$ and $M$, $f(r)=0$ admits zero, one (i.e. two degenerate) or two positive roots. For
\begin{equation}
\ell < \ell_{max}(M)=\sqrt{\frac{16}{27}} M ,   
\end{equation}
the metric possesses two horizons: an outer (event) horizon $r_+$ and an inner (Cauchy) horizon $r_-$, exactly as in the polymerized case discussed above. Instead, at $\ell = \ell_{max}(M)$, the two horizons merge into a single degenerate, extremal horizon, and the BH stops radiating. Finally, for $\ell > \ell_{max}(M)$, no horizon exists and the solution describes a horizon-less compact object. 

The temperature of the Hayward BH is given by:
\begin{equation}
T=\frac{1}{4 \pi r_+}\left(1-\frac{3 \ell^2 }{r_+^2}\right),
\end{equation}
which reduces to the standard Schwarzschild expression $T_{\rm Schw} = 1/(8\pi M)$ for $\ell\to0$, and vanishes as $\ell\to \ell_{max}(M)$.

\subsection{Simpson-Visser regularized solution}

The Simpson-Visser metric \cite{Simpson:2018tsi}, often referred to as the ``black-bounce'' spacetime, is instead the reference example of the second class of RBHs, in which the point-like singularity is replaced by a wormhole throat that remains hidden behind the horizon for an evaporating (i.e. sub-extremal) configuration. The metric reads
\begin{equation}
    ds^2 = -\left(1-\frac{2M}{\sqrt{r^2+\ell^2}}\right)dt^2 + \frac{dr^2}{1-\dfrac{2M}{\sqrt{r^2+\ell^2}}} + (r^2+\ell^2)\,d\Omega^2,
\end{equation}
where the areal radius is $\sqrt{r^2+\ell^2}$ and the radial coordinate $r \in (-\infty,+\infty)$ is allowed to take negative values, with $r=0$ marking the location of the wormhole throat. The parameter $\ell$ plays the role of the regularization scale.

The global structure changes qualitatively with $\ell/M$:
For $0 \le \ell < 2M$, the metric function $f(r) = 1-2M/\sqrt{r^2+\ell^2}$ has a single positive root $r_+ > 0$, so the geometry describes a genuine RBH in which the wormhole throat at $r=0$ is causally hidden behind a single horizon (a ``one-way'', or hidden, wormhole).
At the value $\ell = 2M$, the horizon degenerates onto the throat itself, giving an extremal, horizon-less one-way wormhole. Finally, for $\ell > 2M$, no horizon forms and the geometry is a horizon-less, traversable two-way wormhole.
We again identify
\begin{equation}
\ell_{\rm max}(M) \equiv 2M\,,
\end{equation}
as the transition point, so that $L = \ell/(2M) \in [0,1)$ parametrizes the sub-extremal, evaporating branch.

The temperature of the Simpson Visser BH is given by:
\begin{equation}
T=\frac{\sqrt{4 M^2-\ell^2}}{16 \pi M^2}
\end{equation}
which reduces to the standard Schwarzschild expression $T_{\rm Schw} = 1/(8\pi M)$ for $\ell\to0$, and vanishes as $\ell\to \ell_{max}(M)$.

%======================================================================
\subsection{Results for regularized black holes}
\begin{figure}[h!]
    \centering

    \begin{subfigure}[t]{0.49\linewidth}
        \centering
        \includegraphics[width=\linewidth]
        {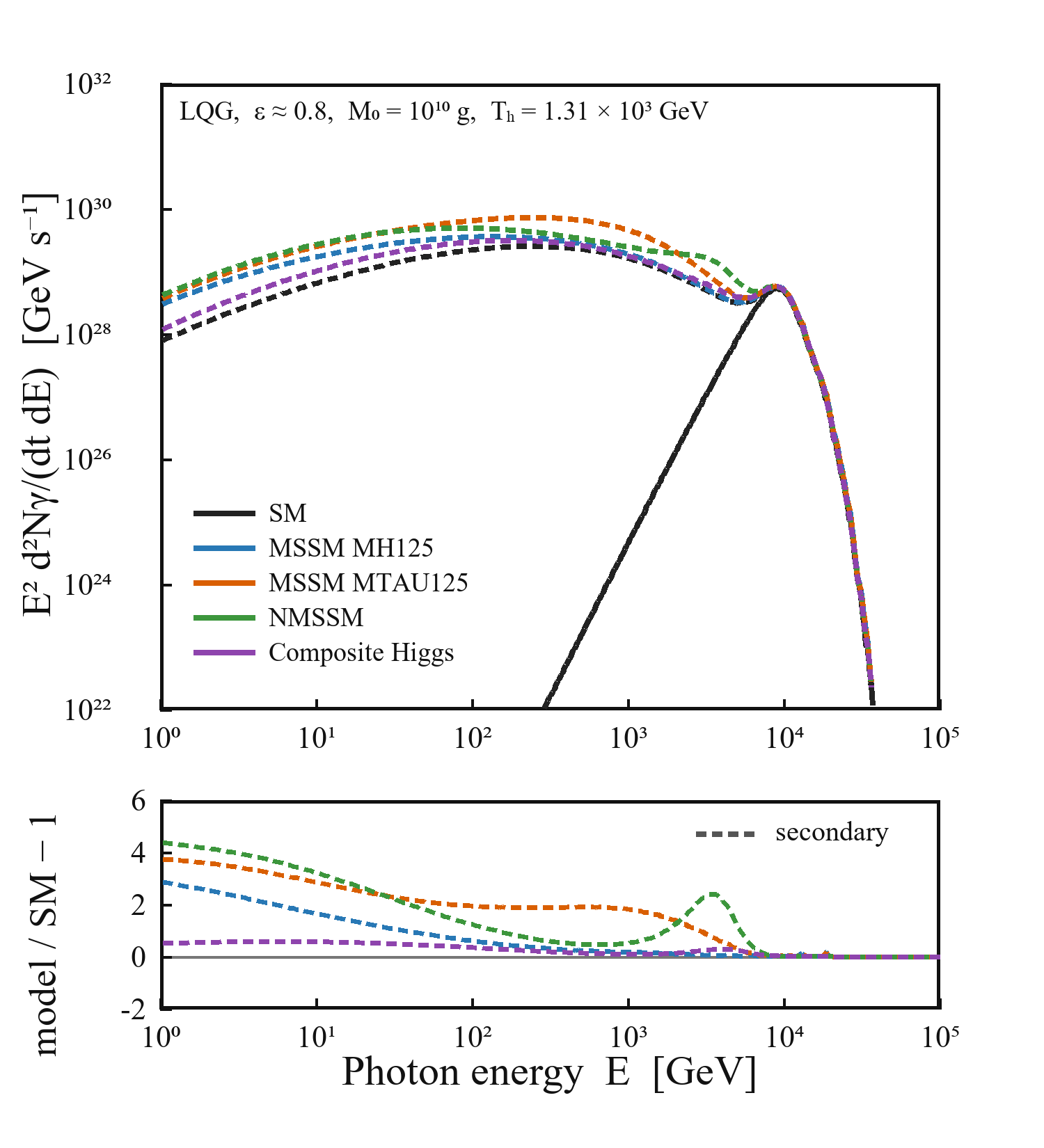}
        \caption{Polymerized, t=0}
        \label{fig:photon_lqg_SuSy_a}
    \end{subfigure}
    \hfill
    \begin{subfigure}[t]{0.50\linewidth}
        \centering
        \includegraphics[width=\linewidth]
        {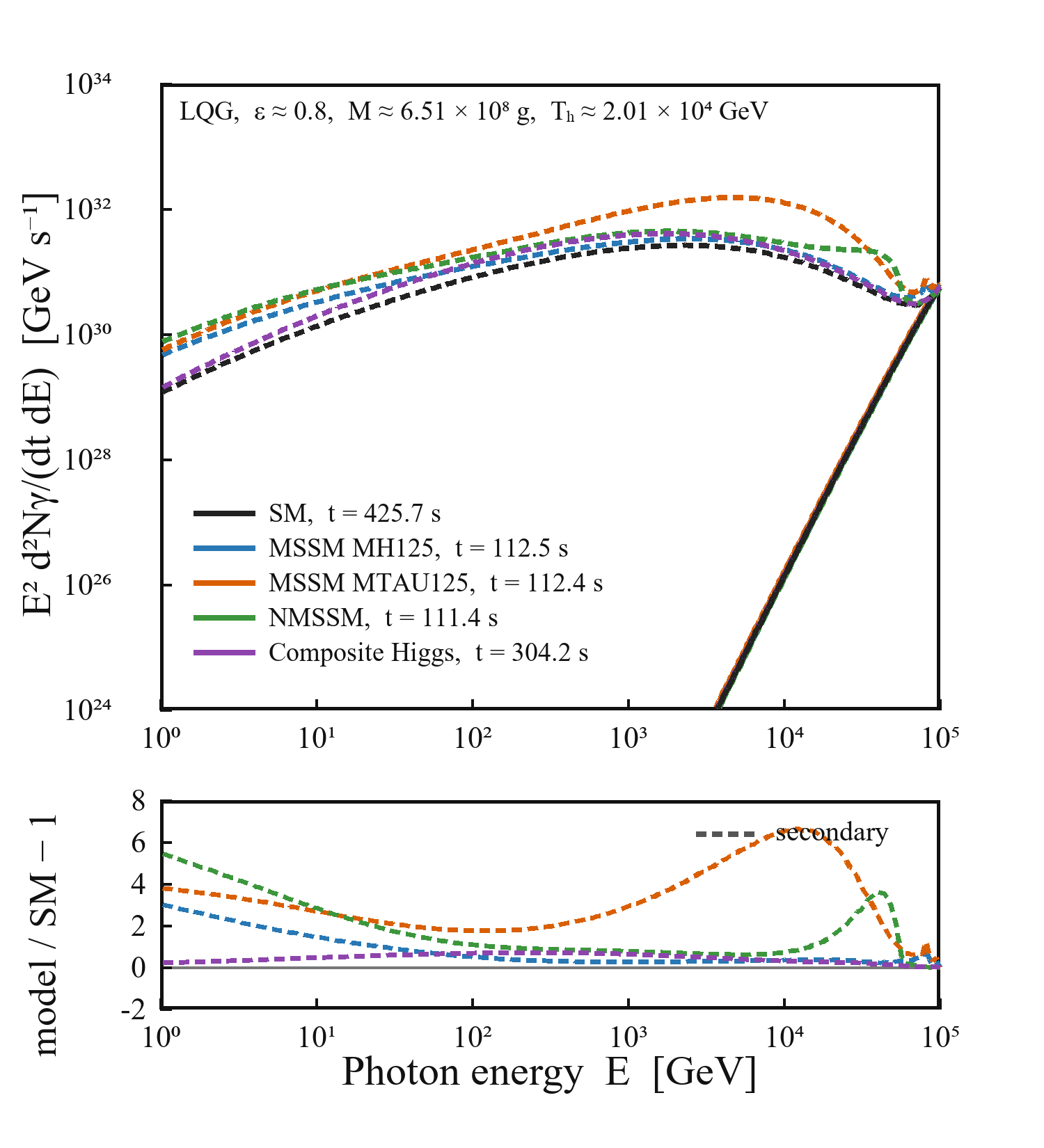}
        \caption{Polymerized, near-final}
        \label{fig:photon_lqg_SuSy_b}
    \end{subfigure}

    \caption{Comparison between the final photon produced in the SM versus SuSy when the evaporating black hole is polymerized.}
    \label{fig:photon_SuSy_lqg}
\end{figure}

Polymerized black holes \cite{polymerized} are characterized by two free parameters, the LQG area gap $a_0$ and the polymerization parameter $\varepsilon$, which are not constrained by the theory. Nevertheless, often $a_0$ is chosen to be \(a_0 \equiv \sqrt{3}\gamma/2 \simeq 0.11\), proportional to the  Barbero--Immirzi parameter $ \gamma \equiv {\ln 2}/{\sqrt{3\pi}} $. In the phenomenological context, only the two values $a_0=0.11$ and  $a_0=0$ are considered. The properties of this solution in the context of Hawking evaporation is extensively studied \cite{Arbey:lqg2,Arbey:potentials} using \texttt{BlackHawk v2.3}, where the authors find that the parameter $a_0$ plays no particular role in the emission rates: a scan over the range between $0$ and $\sqrt{3}\gamma/2$ showed only a very slight decrease at the percentage level in the spectra high-energy tail. Sizeable effects can only be observed when $r_+ \lesssim a_0$, hence only during the very final stages of evaporation of the BH.
Following this result, in the present study we set $a_0=0$. %Another reason for which this choice is justified is that $a_0$, in general, is expected to be negligible for BHs with a radius $r^2_+>>a_0$, and to play a role only at the very final stages of the BH evaporation.  In these conditions, its radius reaches values close to the Planck length, out of the Hawking radiation constraint range.
The parameter $\varepsilon$, on the contrary, has a direct impact on the evaporation spectra, with an effect roughly proportional to its value \cite{Arbey:lqg2}: increasing $\varepsilon$ leads to a decrease in the BH temperature, therefore the overall emission rate also reduces. For $\varepsilon$ between 0 and about $0.8$ (the ``low-$\varepsilon$'' regime), differences with respect to the Schwarzschild case are still modest, vanishing completely if $\varepsilon=0$, as expected. For larger values, roughly between $\varepsilon \gtrsim 1-10$, the impact of LQG corrections becomes more dramatic. 
This reduction is more pronounced for high-spin particles: for a fixed increase in $\varepsilon$, the emission of low-spin particles is less suppressed than that of higher-spin particles. This has clearly direct implications for the relative differences between SM and SuSy/CoDy spectra.  
This effect is evident in the comparison between Figure~\ref{fig:photon_SuSy} and Figure~\ref{fig:photon_SuSy_lqg} with $\varepsilon \simeq 0.8$, where the relative differences between SM and SuSy/CoDy spectra grow when the parameter increases from 0 (Schwarzschild limit). This interesting feature can be explained with the following considerations: the dominant channel for photon production in the SM is $\pi^0 \rightarrow \gamma \gamma$ with branching ratio $\mathrm{BR} \approx 0.99$. Neutral pions are mainly produced by quarks (spin-$1/2$) and, in the SuSy extensions, by squarks (spin-0), since the main decay channel for squarks (from the \texttt{SPheno} decay tables) is into quarks. Therefore, as $\varepsilon$ increases, quark emission in the SM is more strongly suppressed than squark emission in SuSy, due to the spin dependence of the potentials. As a result, the secondary photon spectrum in the SM is more reduced, which makes their relative difference more pronounced.

The other two RBH solutions depend instead on the length scale $\ell$: its impact on evaporation can be inferred by comparing it to a characteristic function of the BH mass, $\ell_\mathrm{max}(M)$, which depends on the specific metric, as described in the previous sections.  The physical scale $\ell$, instead, must be considered a constant of the theory, as it is a spacetime parameter that fixes the scale at which new gravitational physics effects become important. The effect on evaporation is then encoded in a dimensionless parameter $L$, see Eq.~\eqref{eq:ratio}, inversely proportional to the mass $M$. As a consequence, as the BH evaporates and loses mass, $L$ increases until the BH becomes extremal ($L=1$).
If $\ell$ is assumed to be fixed at the Planck length, $\ell=\ell_{\rm P}$, sizeable effects (i.e. large $L$), only occur for BH masses close to the Planck one.

To see when gravitational effects become comparable to the BSM effects at the TeV scale, we fixed the physical BH mass and
determined the regularization length $\ell$ required to obtain a given value of $L$. Considering a BH with a reference initial mass of $M_{\rm BH}=10^{10}\,\mathrm{g}$, the corresponding geometrical mass scale is $M = \frac{GM_{\rm BH}}{c^2} \simeq 7.43\times10^{-21}\,\mathrm{m}$,
and its Schwarzschild radius is $r_{\rm S}=2M
\simeq
1.49\times10^{-20}\,\mathrm{m}$. 
For the Simpson--Visser metric, the physical regularization length in terms of the Schwarzchild radius is
\begin{equation}
\ell_{\rm SV}
=
2ML_{\rm SV}
=
r_{\rm S}L_{\rm SV}.
\end{equation}
Similarly, for the Hayward metric we have
\begin{equation}
\ell_{\rm H}
=
\sqrt{\frac{16}{27}}\,
ML_{\rm H}
=
\frac{4}{3\sqrt{3}}\,
ML_{\rm H}.
\end{equation}
Some values of the regularized physical scale for fixed values of $L$ are summarized in Table~\ref{tab:regularization_length_1e10g}.

\begin{table}[h]
\centering
\begin{tabular}{ccc}
\hline
Metric
&
Normalized parameter
&
$\ell$ [m]
\\
\hline
Simpson--Visser
&
$L_{\rm SV}=0.7$
&
$1.04\times10^{-20}$
\\

Simpson--Visser
&
$L_{\rm SV}=0.8$
&
$1.19\times10^{-20}$
\\

Hayward
&
$L_{\rm H}=0.7$
&
$4.00\times10^{-21}$
\\

Hayward
&
$L_{\rm H}=0.8$
&
$4.57\times10^{-21}$
\\
\hline
\end{tabular}
\caption{Physical regularization lengths required to obtain normalized
parameters between \(0.7\) and \(0.8\) for a black hole with
\(M_{\rm BH}=10^{10}\,\mathrm{g}\).}
\label{tab:regularization_length_1e10g}
\end{table}

Hence, for a BH with mass \(10^{10}\,\mathrm{g}\) and Hawking temperature $T_H \sim 1$~TeV, a normalized parameter $L$
of order \(0.7\)--\(0.8\) requires regularization scales of the order of
the BH geometrical radius $\sim 10^{-20}$ m, which is much larger than the Plank length.
Therefore, these choices do not correspond to a
Planck-scale correction, but describe a strong geometrical
deformation acting on a scale comparable to the horizon size. Whith this choice of $\ell$, combined effects of BSM physics and geometrical correction can intertwine. 
We use as a reference model the $M_h^{125}$ SuSy benchmark, keeping in mind that all the considerations above about the effects of BSM particles still apply. 

\begin{figure}[h!]
    \centering   \includegraphics[width=1.05\linewidth]{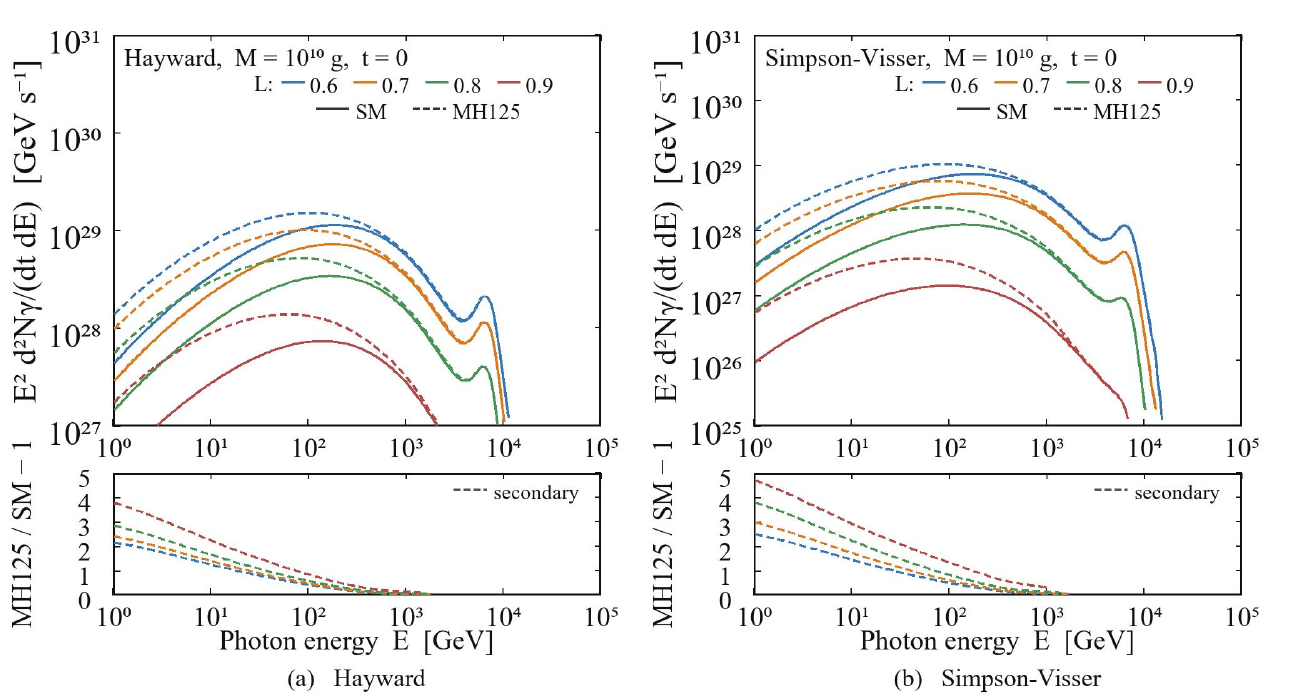}
    \caption{Comparison between Hayward and Simpson--Visser metric, for a black hole of mass $10^{10}$g and a scan of different regularization parameters $L$.}
    \label{fig:SuSy_reg}
\end{figure}

The results are shown in Figure~\ref{fig:SuSy_reg} for Hayward (left) and Simpson-Visser (right), for different values of $L$. As expected from previous studies, the RBH metrics typically reduce the flux of secondary particles in the low energy part of the spectrum, here demonstrated for photons. We also include the effect of BSM physics, via the SuSy benchmark $M_h^{125}$, where the bottom panels indicate the ratio to the corresponding SM spectra. We see that in the BSM scenario, the low energy spectrum is strongly enhanced. The main emission of secondary photons in those tails is due to the primary emission of scalar squarks. Similar effects can be expected in the other SuSy cases.

%====================================================================================
\section{Comparison with observational ranges}
\label{sec:detection}

The results presented in the previous sections show that the additional degrees
of freedom of the BSM scenarios considered here produce an excess of secondary
photons with respect to the SM, whose size and shape depend on the
benchmark. What matters for observation is not the size of the excess but its
location: the peaks of the relative enhancement consistently lie in the range
$\sim10^{2}$--$10^{4}$~GeV, which overlaps the sensitivity windows of current
gamma-ray instruments. Fermi--LAT \cite{FermiLAT:2018PBH} covers the lower end of this interval up to about $1$~TeV, while Imaging
Atmospheric Cherenkov Telescopes such as HESS \cite{HESS_Hinton_2004,
HESS:2021rto} and wide-field air shower arrays such as HAWC
\cite{HAWC:2013kzm, HAWC_Abeysekara_2013} and LHAASO
\cite{LHAASO_cao2022largehighaltitudeair} are sensitive in the multi-TeV domain (C.f., Table~\ref{tab:instruments}). The predicted enhancement is
not only a theoretical curiosity but falls inside the detectable window of
present and next-generation experiments. Should photon emission from an
evaporating BH ever be observed \cite{Ukwatta:2015iba}, the data would
offer a test of BSM physics complementary to collider
searches at the LHC \cite{LHC_Evans:2008zzb, ATLAS:2022ihe, CMS:2019pov} and at
future accelerator facilities.

\renewcommand{\arraystretch}{1.3}
\begin{table}[h!]
\centering
\vspace{0.2cm}
\begin{tabular}{lcc}
\hline\hline
Instrument & Technique & Energy range \\
\hline
Fermi--LAT & Space-based pair-conversion telescope 
           & 20 MeV -- $300$ GeV \\
\hline
H.E.S.S. & \makecell{Imaging Atmospheric Cherenkov \\ Telescopes (IACTs)}
          & 50 GeV -- 100 TeV \\
\hline
HAWC & Wide-field air-shower array 
     & 100 GeV -- 100 TeV \\
\hline
LHAASO & Hybrid wide-field air-shower observatory 
       & $\sim 100$ GeV -- PeV \\
\hline\hline
\end{tabular}
\caption{Energy coverage of the main $\gamma$-ray instruments relevant for
this study.}
\label{tab:instruments}
\end{table}

Neutrino fluxes are also affected at energy scales around the TeV, within
reach of neutrino telescopes such as IceCube \cite{gaisser2014icecube}, ANTARES
\cite{amram2002antares} and KM3NeT \cite{km3net, km3net2025observation}. Compared to
photon detectors, neutrino observatories are strongly sensitive to the incoming
flux intensity, due to the weaker interactions of neutrinos to matter, and to the exposure time, which is a severe limitation here: a
black hole of mass $M=10^{10}$\,g evaporates completely within a few minutes, so
the signal from its final stages is correspondingly brief. 
The flux intensity in neutrinos also remains comparable to that into photons, which, therefore,
remain the most promising channel for testing the underlying BSM model.

\begin{figure}[h!]
    \centering
    \includegraphics[width=0.6\linewidth]{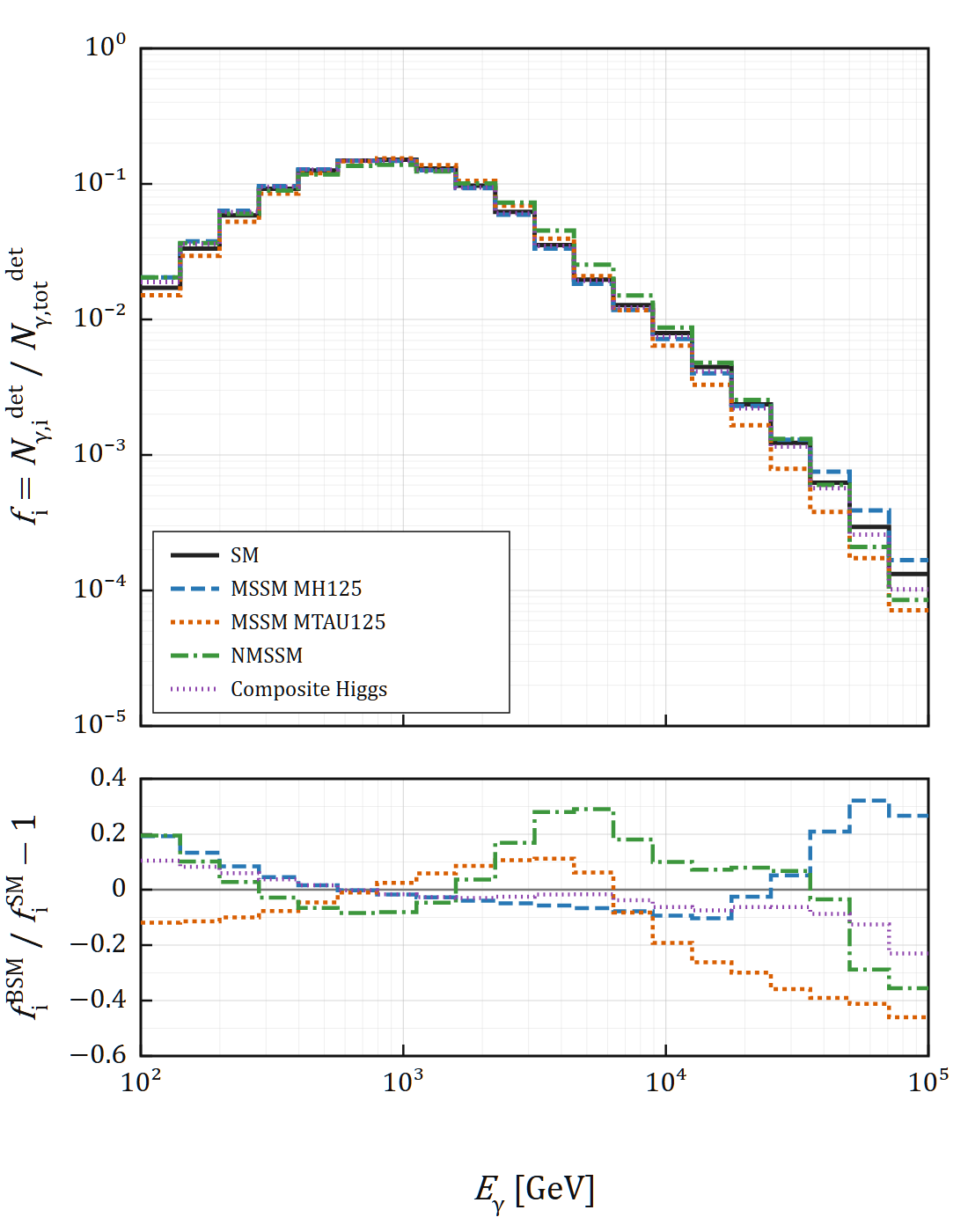}
\caption{Normalized HAWC photon-event spectra for a Schwarzschild black hole with initial mass \(M=10^{10}\,\mathrm{g}\), comparing the SM with the MSSM, NMSSM, and Composite Higgs scenarios. The secondary photon spectra are integrated over the black-hole evaporation history, folded with the HAWC effective area, and divided into 20 logarithmic energy bins between \(10^{2}\) and \(10^{5}\,\mathrm{GeV}\). The upper panel shows the fraction of detected photons in each bin, \(f_i=N_{\gamma,i}^{\mathrm{det}}/N_{\gamma,\mathrm{tot}}^{\mathrm{det}}\). The lower panel shows the relative deviation of each BSM prediction from the SM result, \(f_i^{\mathrm{BSM}}/f_i^{\mathrm{SM}}-1\).}
    \label{fig:spectra_hawk}
\end{figure}

To illustrate the level of sensitivity a gamma ray detector must have to distinguish among different BSM scenarios, we considered the case of HAWC \cite{HAWC:2013kzm}. Within the sensitivity range between $100$~GeV and $100$~TeV, see Table~\ref{tab:instruments}, we computed the number of photons in energy bins normalized to the total number of emitted photons for a SM emission and for each BSM benchmark, defined as:
\begin{equation}
f_j
=
\frac{
\displaystyle
\int_{E_j}^{E_{j+1}} dE\,
A_{\rm eff}(E)
\int_{t_0}^{t_f} dt\,
\frac{d^2N_\gamma}{dE\,dt}(E,M(t))
}{
\displaystyle
\int_{E_{\min}}^{E_{\max}} dE\,
A_{\rm eff}(E)
\int_{t_0}^{t_f} dt\,
\frac{d^2N_\gamma}{dE\,dt}(E,M(t))
}
\end{equation}
where \(A_{\rm eff}(E)\) denotes the detector effective area for events arriving within the \(0^\circ\!-\!26^\circ\) zenith-angle range, derived from \cite{HAWC:2024plu}. The results are shown in Figure~\ref{fig:spectra_hawk}, where we considered the total emission for a BH of initial mass $M = 10^{10}$~g.

For this comparison, we assume that the detector measures the total emission over the BH evaporation, due to the short lifetime of the considered mass. We see in Figure~\ref{fig:spectra_hawk} that each BSM benchmark produce characteristic enhancements or suppression in different energy bins, as it is clearly showed in the bottom panel. Hence, comparing the measured flux with specific templates may allow to distinguish the BSM benchmarks from the SM emission. As we normalized the flux in each bin as a fraction of the total photons in the whole energy range, the comparison does not depends on the distance of the source from the detector, nor on changes in the total emitted photons due to the shortening of the BH lifetime in BSM scenarios with respect to the SM case.

%======================================================================
\section{Conclusions and outlook}\label{sec:concl}
We have studied how physics beyond the Standard Model, both in the
particle sector and in the black hole metric, affects the high-energy photon and
neutrino fluxes emitted by evaporating black holes. On the particle
side, we extended the standard evaporation treatment to three supersymmetric
benchmarks --- two within the MSSM ($M_h^{125}$, $M_\tau^{125}$) and one within the NMSSM --- and to a composite
Higgs scenario. We computed numerical tables of the transfer functions allowing to compute the secondary particle fluxes for all benchmarks, employing \texttt{Pythia} with a uniform
particle-gun setup. On the gravity side, we considered three
regular metrics implemented in \texttt{BlackHawk v3.0}: a polymerized Schwarzschild solution inspired by loop quantum
gravity, the Hayward metric representative of a regular-core solution, and the Simpson--Visser metric illustrating hidden-throat classes. Numerically, we set the scale of new particle masses in the BSM benchmarks to lie close to the TeV scale, so that the Hawking radiation is affected for BH masses smaller than roughly $10^{10}$~g, corresponding to Hawking temperatures above $1$~TeV. Correspondingly, we chose  the parameters of the regular BH metrics affect the emission at the same BH mass scales.

The spectra of photons and neutrinos, produced via \texttt{BlackHawk v3.0} combined with our transfer functions, show that BSM scenarios produce excesses at energies close to the TeV, within range of present instruments. The shape of the excess
discriminates between benchmarks: a direct radiative decay produces a narrow bump
whose height is stable with temperature, as in the NMSSM through the neutralino decay
$\widetilde\chi^0_2\to\widetilde\chi^0_1\gamma$, while heavy states feeding
multistep cascades produce a broad enhancement that grows with temperature, as in
$M_\tau^{125}$; a compressed electroweak spectrum, as in $M_h^{125}$, gives no
localized feature at all. The composite Higgs benchmark shows a narrow feature of the
first kind, from the decay of a singlet pion-like state $s\to Z\gamma$. We also showed that BH geometry and particle content do not
decouple: because the regularized metrics suppress emission more strongly for
higher-spin states, the relative difference between the Standard Model and its
extensions is itself sensitive to the deformation parameter, so the two classes
of new physics can be probed in the same observable.

Several directions remain open. If photons emitted by an evaporating primordial black hole, or a distant localized population of black holes, is observed, the instantaneous spectra we computed should be confronted to the observed signals integrated over the instrument exposure time. As an example, we provided a simple illustration for the typical parameters of the HAWC gamma ray telescope. The neutrino channel is limited here by flux and exposure time, nevertheless it may be interesting to explore for the emission of very energetic neutrinos during the final phases of evaporation, and for a population rather than a single object. We provided numerical tables that can make such data analyses streamlined, once incorporated in public codes such as \texttt{BlackHawk}. On the gravity side, we only considered parameters that give effects for black hole masses of $10^{10}$~g, in order to compare them to the BSM scenarios at the TeV scale. This corresponds to modifications of gravity at the scale of the radius, $\ell\sim 10^{-20}$~m, rather than at the Planck length. Whether a physically motivated choice of $\ell$ can still leave an observable imprint deserves a dedicated study.

\section*{Acknowledgements}
We thank Abhishek Singh for his assistance in implementing the Hidden Valley module of \texttt{Pythia} for the CoDy numerical computations.  We also thank Jessica Turner for useful discussions.
The work of F.S. and V.V. is partially supported by the Carlsberg Foundation, \emph{semper ardens} grant CF22-0922.
G.C. acknowledges support from the project Tremplin$@$Physique 2025, funded by the Institute of Physics (INP) of the Centre National de la Recherche Scientifique (CNRS) of France.

\section*{Data availability}
All data has been generated by using the public available codes \texttt{BlackHawk v3.0} and \texttt{Pythia v8.3}. The tables with the transfer functions are available at \href{https://github.com/PaolaSimone0/tables_transfer_functions}{this} GitHub repository.

\section*{AI usage}
The authors used AI assistance for language editing, reviewed the resulting text, and take full responsibility for the final manuscript.

\newpage
\appendix
\section{Additional secondary spectra.}\label{sec:appendix}
\begin{figure}[h!]
    \centering

    \begin{subfigure}[t]{0.49\linewidth}
        \centering
        \includegraphics[width=\linewidth]
        {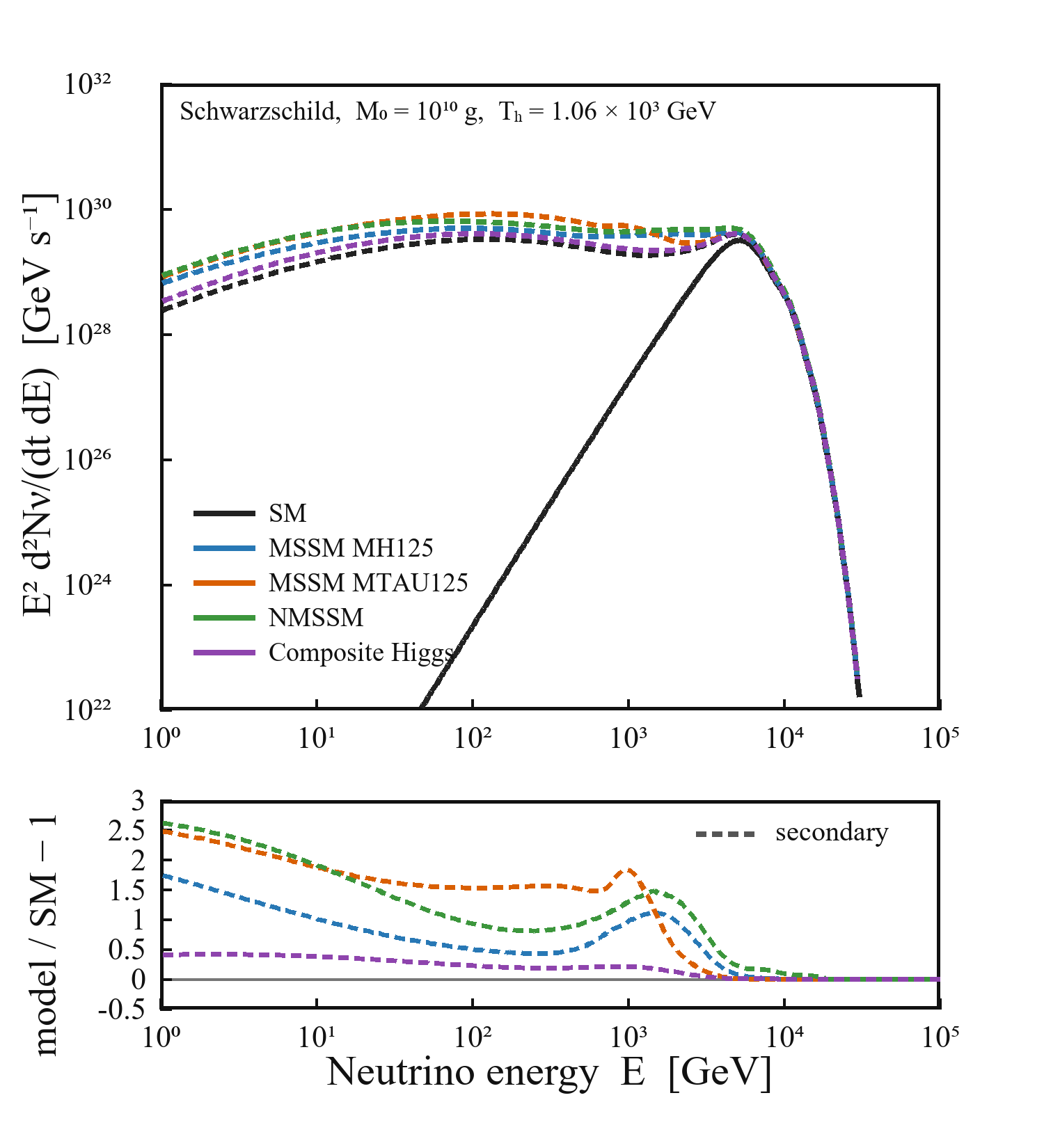}
        \caption{Schwarzschild, t=0}
   \label{fig:neutrino_SuSy_initial}
    \end{subfigure}
    \hfill
    \begin{subfigure}[t]{0.50\linewidth}
        \centering
        \includegraphics[width=\linewidth]
        {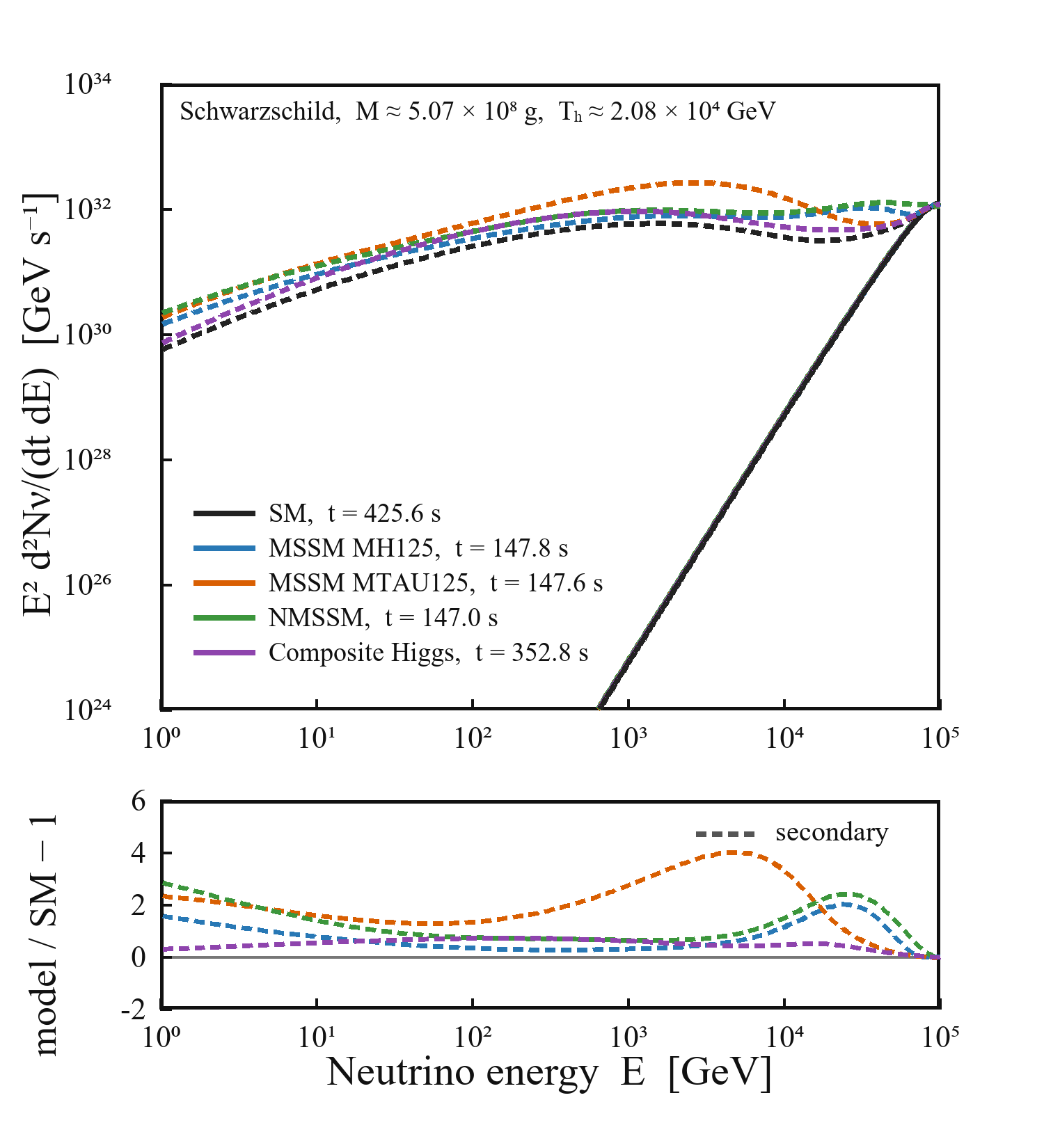}
        \caption{Schwarzschild, near-final}
         \label{fig:neutrino_SuSy_final}
    \end{subfigure}

    \caption{Time evolution of the secondary neutrino spectrum for a
     Schwarzschild black hole of initial mass $M=10^{10}$\,g, comparing across the different benchmarks, with the ratio of
     Eq.~(\ref{eq:ratio}) in the lower panels.}
    \label{fig:neutrino_SuSy_ch}
\end{figure}

\begin{figure}[h!]
    \centering

    \begin{subfigure}[t]{0.49\linewidth}
        \centering
        \includegraphics[width=\linewidth]
        {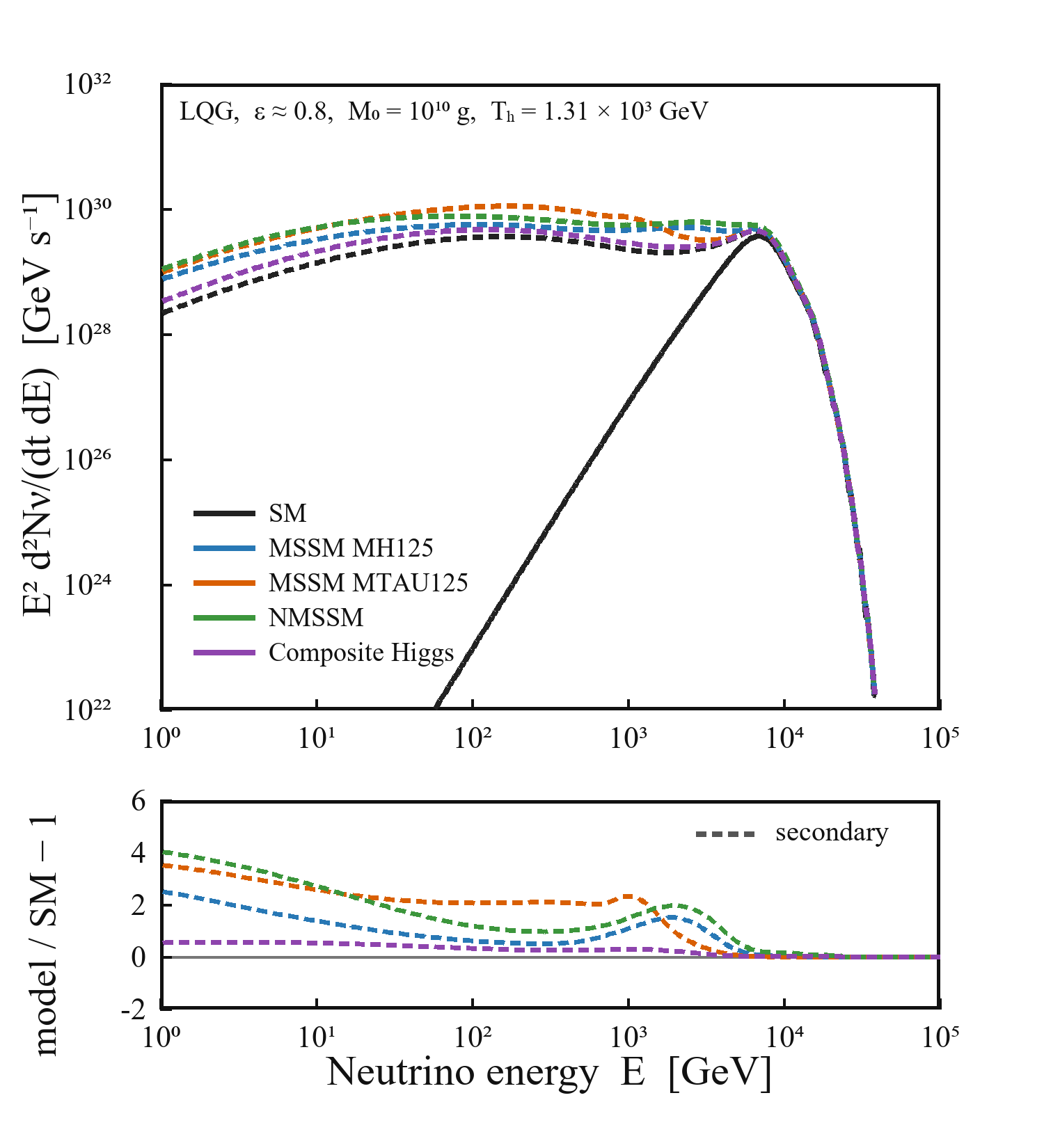}
        \caption{Polymerized, t=0}
   \label{fig:lqg_neutrino_SuSy_initial}
    \end{subfigure}
    \hfill
    \begin{subfigure}[t]{0.50\linewidth}
        \centering
        \includegraphics[width=\linewidth]
        {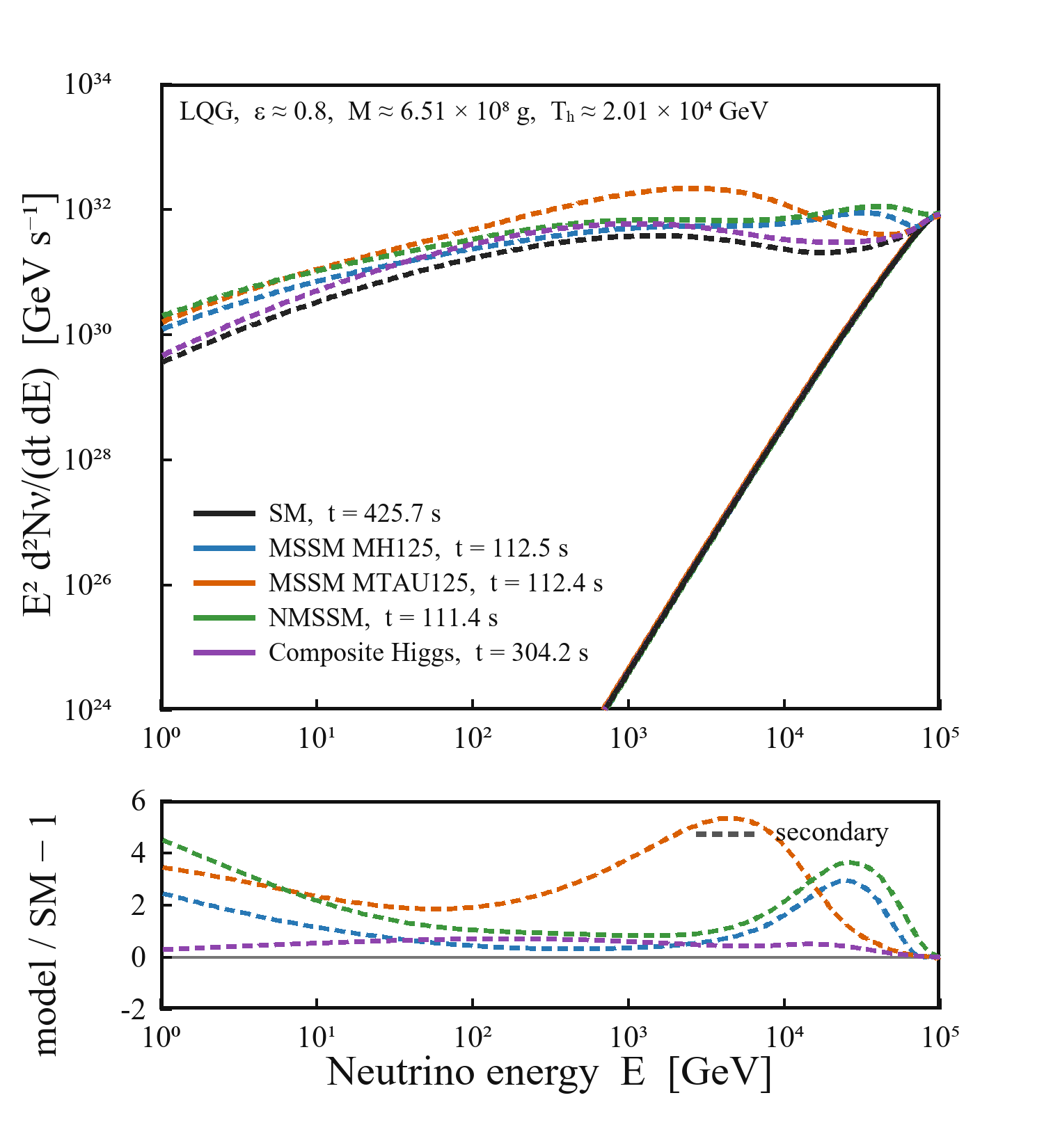}
        \caption{Polymerized, near-final}
         \label{fig:lqg_neutrino_SuSy_final}
    \end{subfigure}

    \caption{Time evolution of the secondary neutrino spectrum for a
     black hole of initial mass $M=10^{10}$\,g, and a near-final stage of temperature $T_H$= 20 TeV.}
    \label{fig:lqg_neutrino_SuSy_ch}
\end{figure}

\begin{figure}[htbp]
  \centering

  % --- first row ---
  \begin{subfigure}{0.47\textwidth}
    \centering
    \includegraphics[width=\linewidth]
      {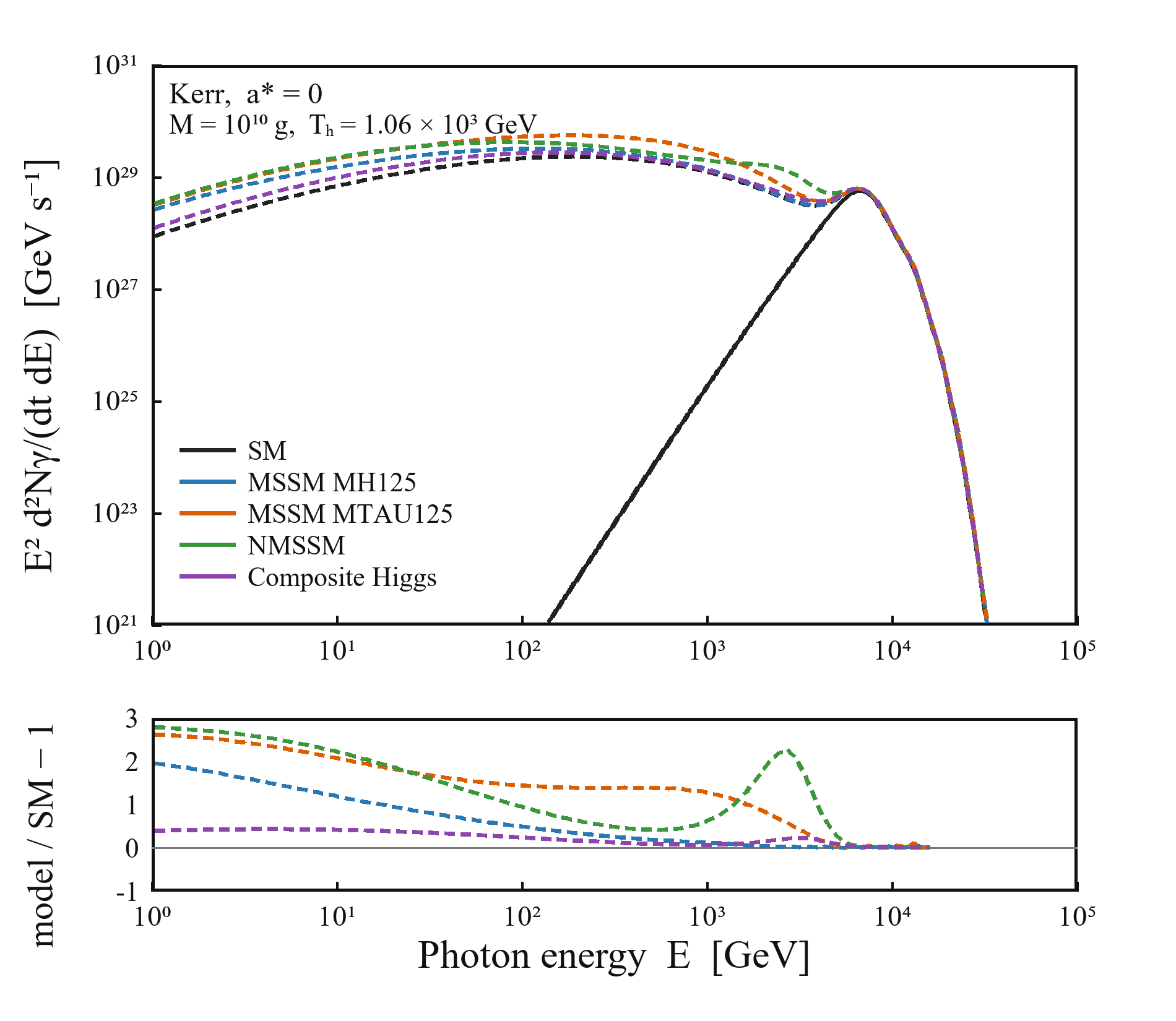}
    \caption{$a_\ast = 0$}
    \label{fig:photon-models-kerr-a0}
  \end{subfigure}
  \hfill
  \begin{subfigure}{0.47\textwidth}
    \centering
    \includegraphics[width=\linewidth]
      {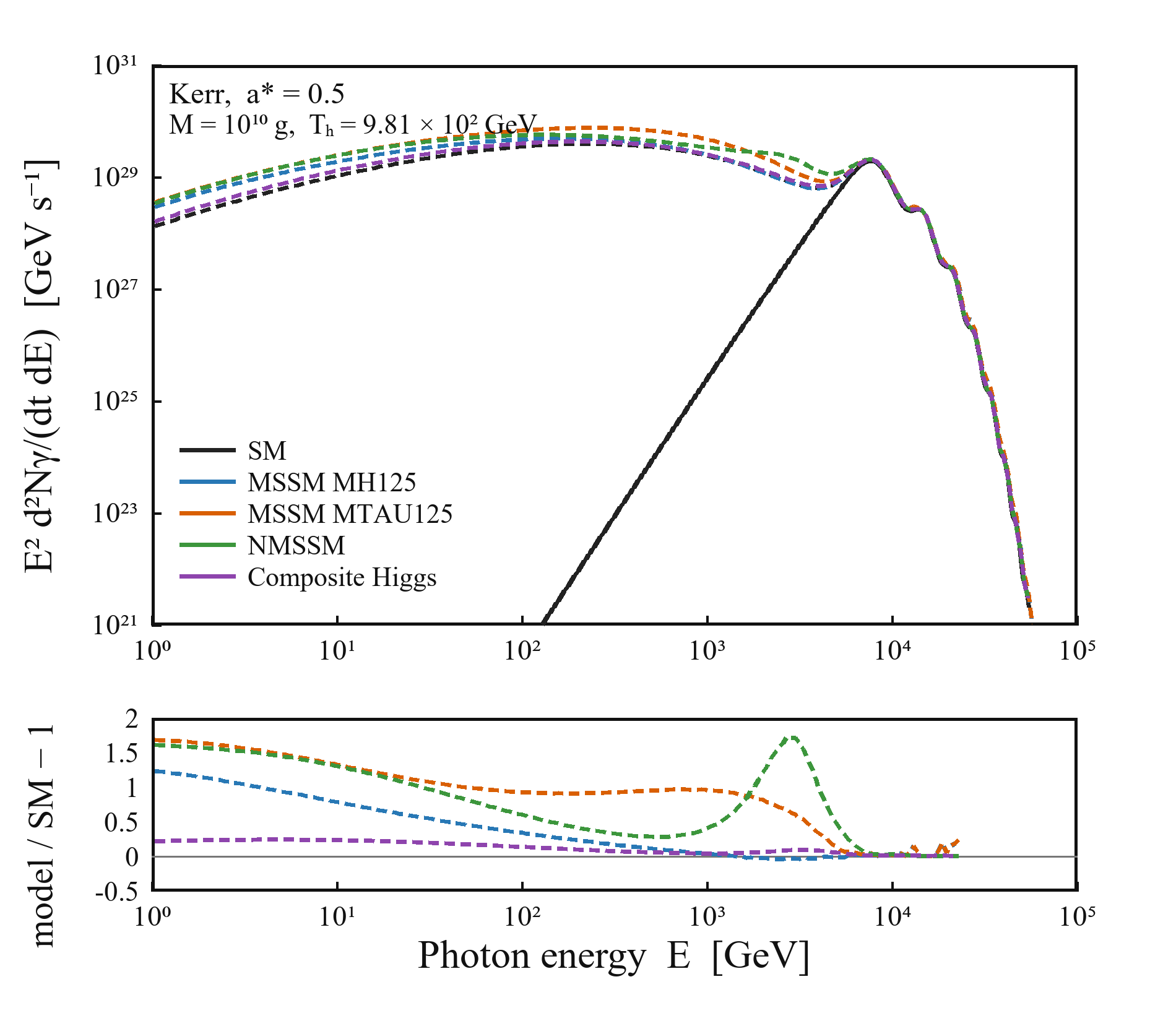}
    \caption{$a_\ast = 0.5$}
    \label{fig:photon-models-kerr-a05}
  \end{subfigure}

  \vspace{0.4cm}

  % --- second row ---
  \begin{subfigure}{0.47\textwidth}
    \centering
    \includegraphics[width=\linewidth]
      {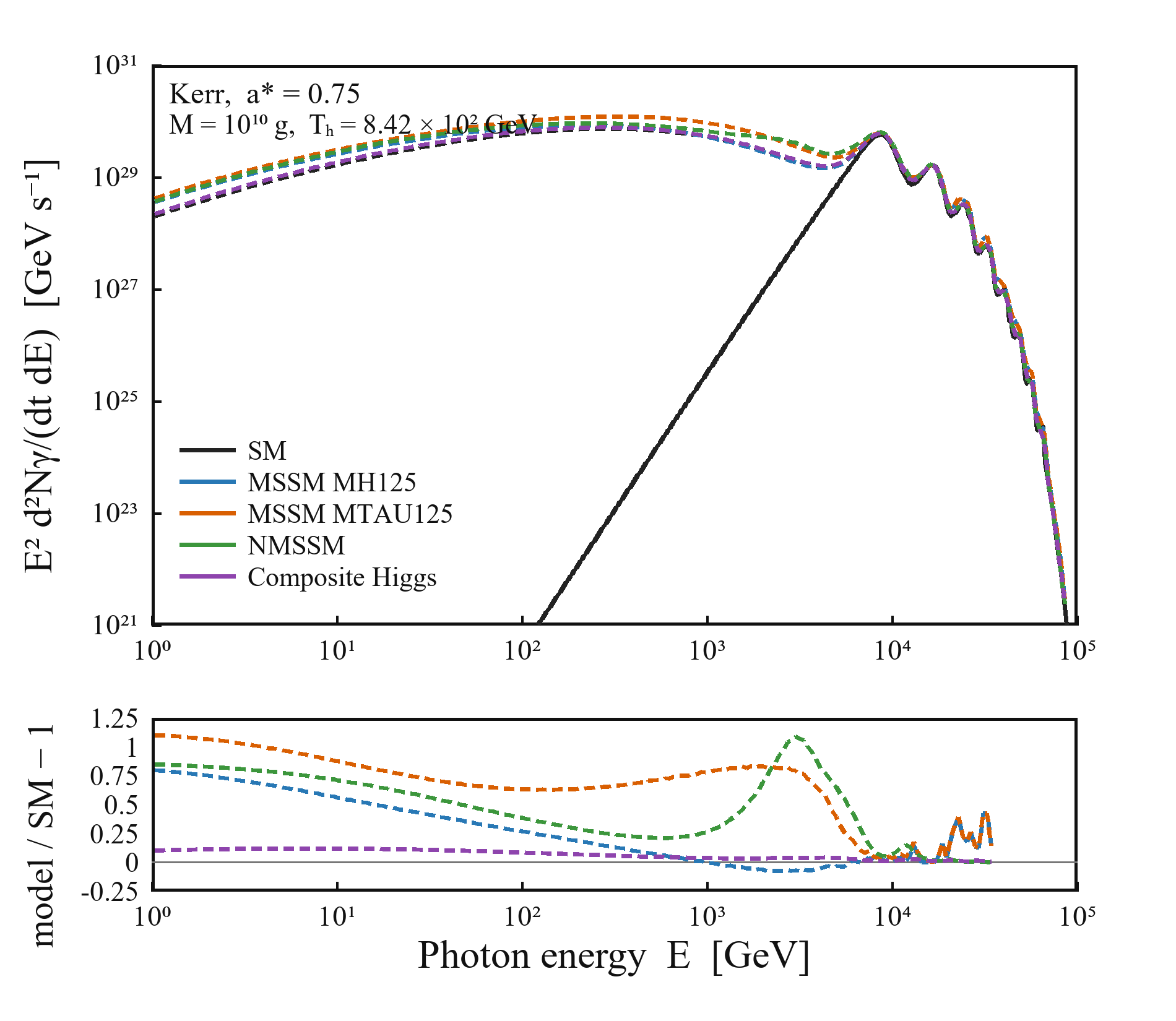}
    \caption{$a_\ast = 0.75$}
    \label{fig:photon-models-kerr-a075}
  \end{subfigure}
  \hfill
  \begin{subfigure}{0.47\textwidth}
    \centering
    \includegraphics[width=\linewidth]
      {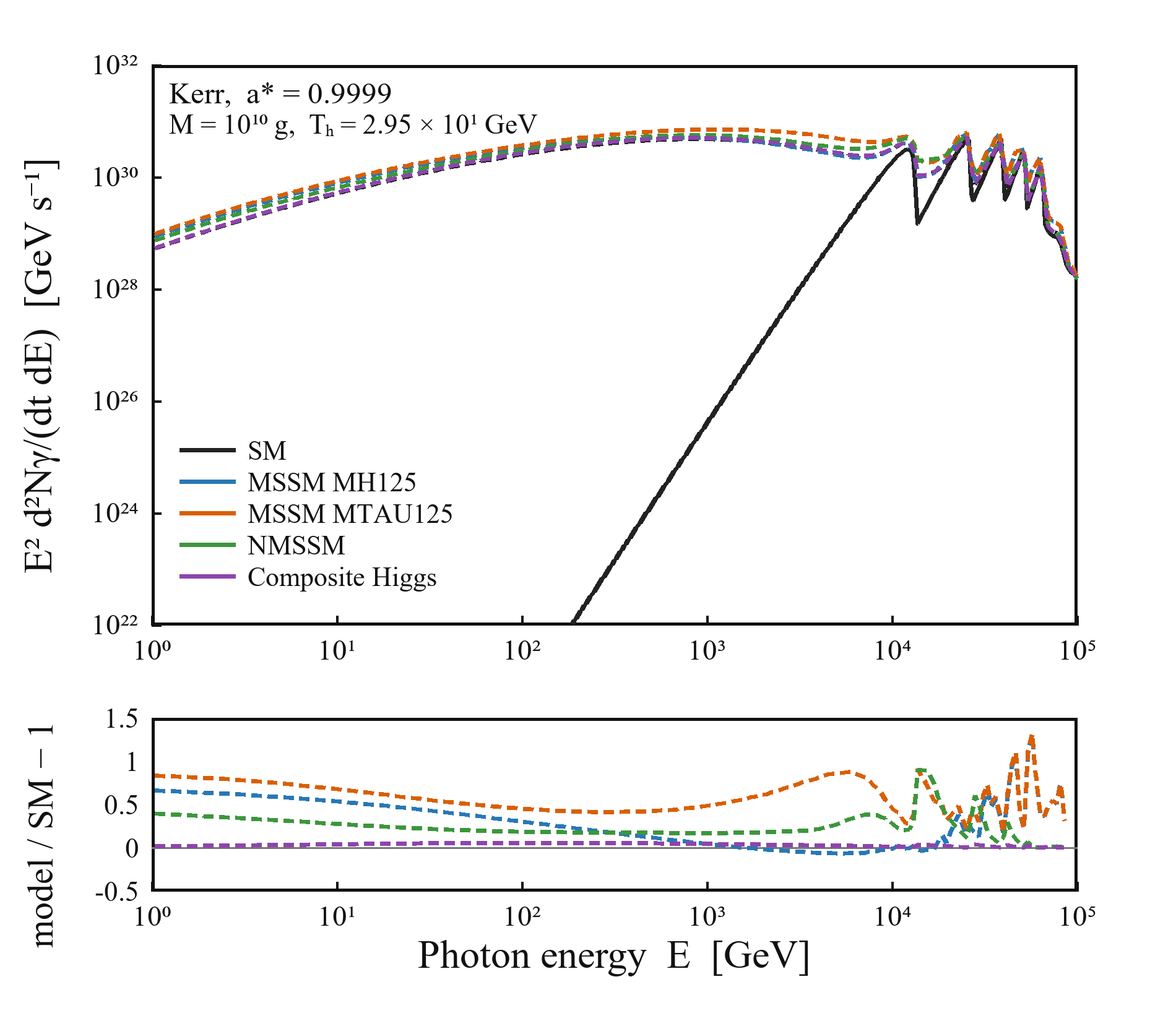}
    \caption{$a_\ast = 0.9999$}
    \label{fig:photon-models-kerr-a09999}
  \end{subfigure}

  \caption{Instantaneous photon emission spectra for Kerr black holes
  with fixed mass $M=10^{10}\,\mathrm{g}$ and different dimensionless
  spin parameters $a_\ast$. Solid and
  dashed lines represent the primary and secondary photon components,
  respectively.}
  \label{fig:photon-particle-models-kerr}
\end{figure}

\begin{figure}[htbp]
  \centering

  % --- first row ---
  \begin{subfigure}{0.47\textwidth}
    \centering
    \includegraphics[width=\linewidth]
      {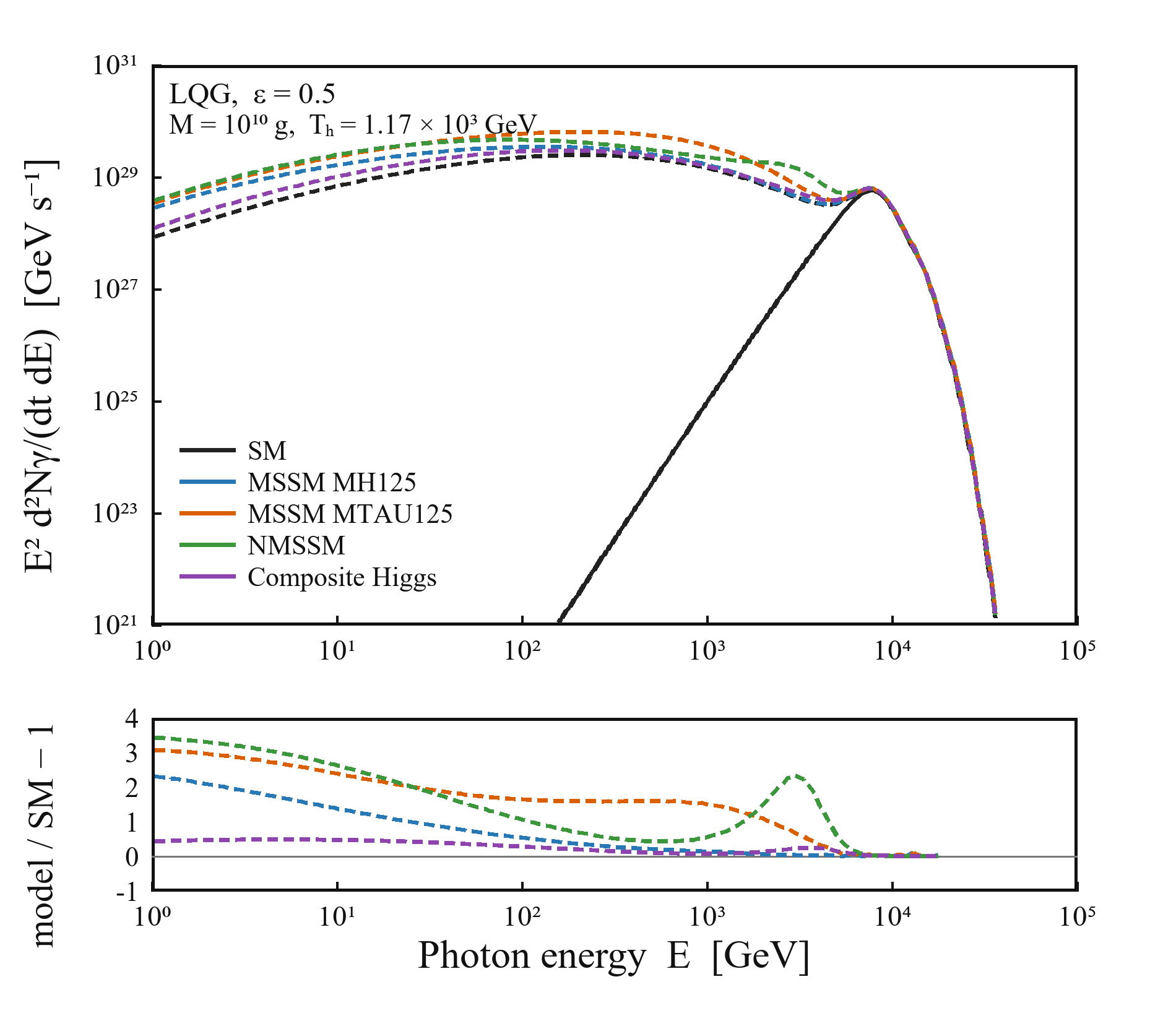}
    \caption{$\varepsilon = 0.5$}
    \label{fig:photon-models-lqg-eps05}
  \end{subfigure}
  \hfill
  \begin{subfigure}{0.47\textwidth}
    \centering
    \includegraphics[width=\linewidth]
      {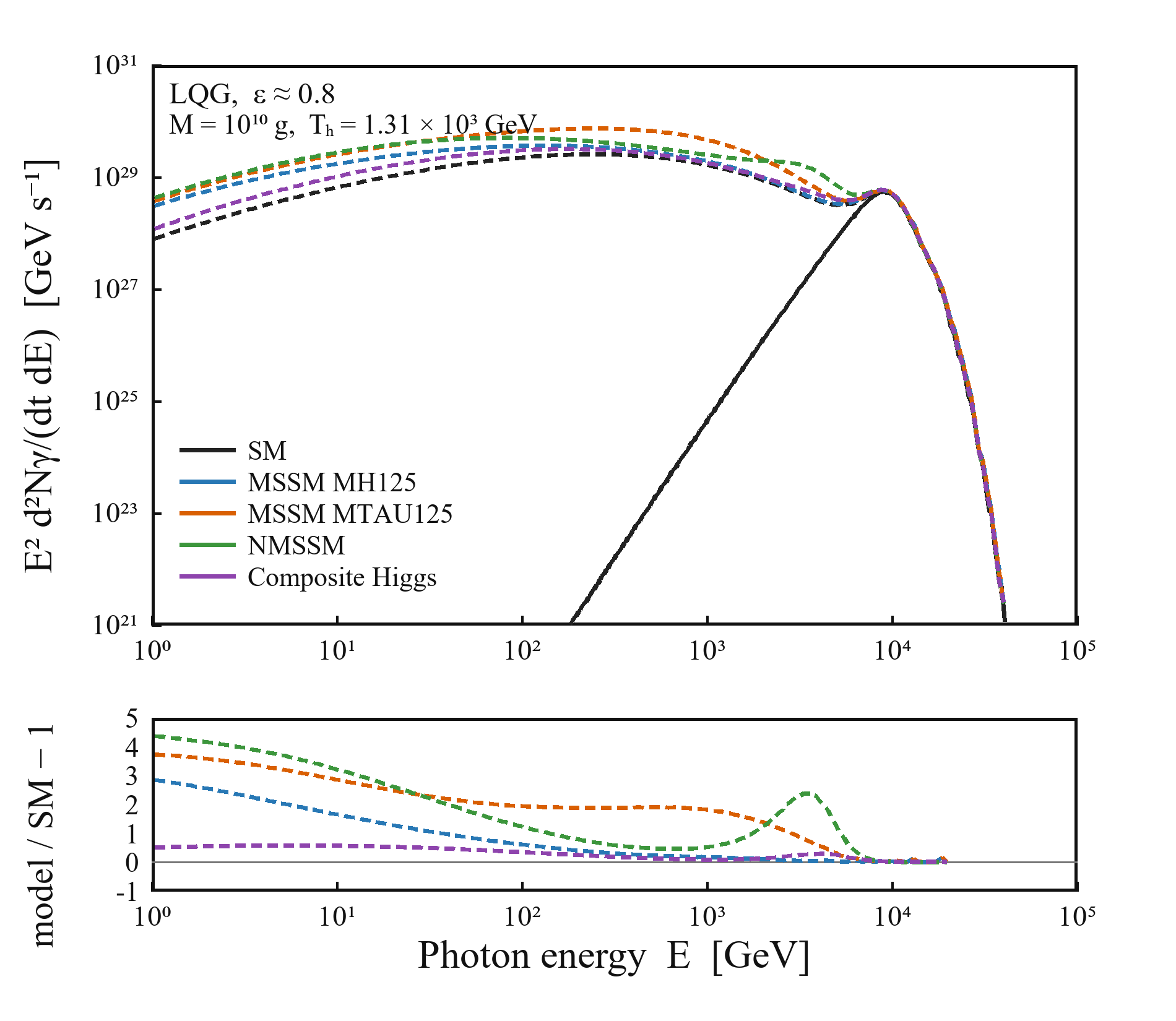}
    \caption{$\varepsilon \simeq 0.8 $}
    \label{fig:photon-models-lqg-eps0794328}
  \end{subfigure}

  \vspace{0.4cm}

  % --- second row ---
  \begin{subfigure}{0.47\textwidth}
    \centering
    \includegraphics[width=\linewidth]
      {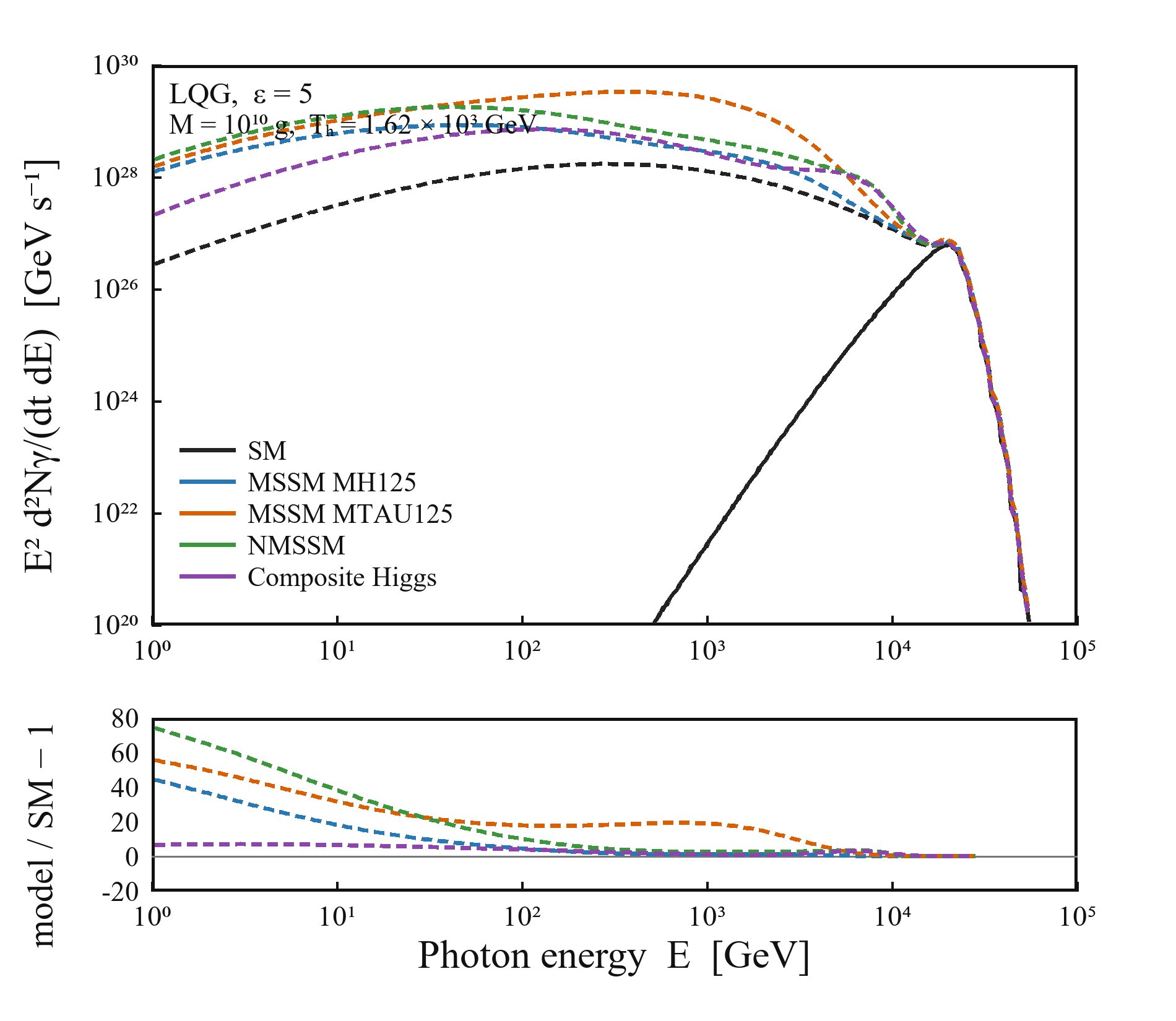}
    \caption{$\varepsilon = 5$}
    \label{fig:photon-models-lqg-eps5}
  \end{subfigure}
  \hfill
  \begin{subfigure}{0.47\textwidth}
    \centering
    \includegraphics[width=\linewidth]
      {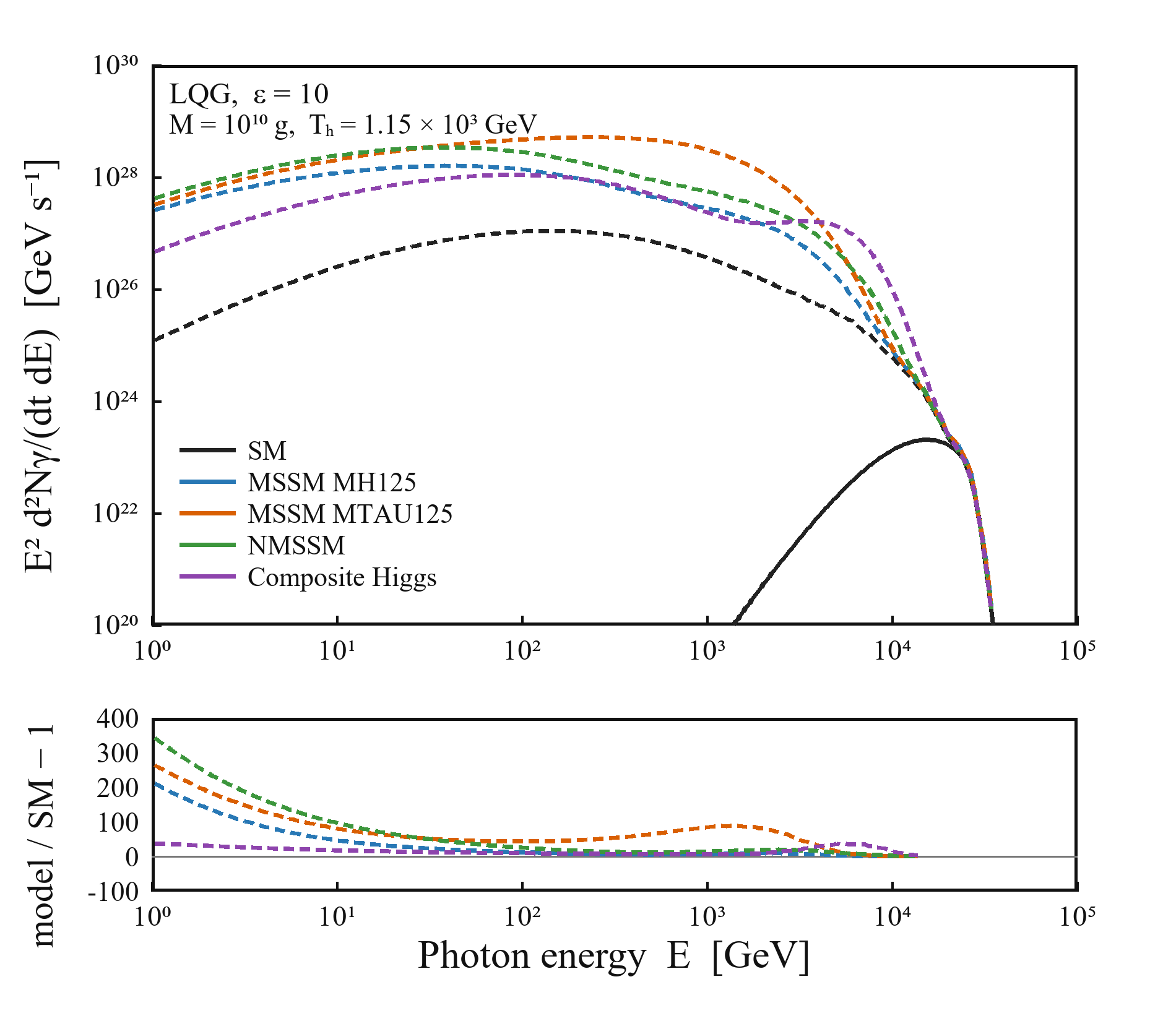}
    \caption{$\varepsilon = 10$}
    \label{fig:photon-models-lqg-eps10}
  \end{subfigure}

  \caption{Instantaneous photon emission spectra for polymerized LQG
  black holes with fixed mass $M=10^{10}\,\mathrm{g}$ for
  different physical values of the polymerization parameter $\varepsilon$. While the semiclassical regime naturally suggests $\varepsilon<1$, we also explore larger values of the polymerization parameter, following previous phenomenological studies in which no a priori upper bound on $\varepsilon$ is imposed \cite{Arbey:lqg2}}
  \label{fig:photon-particle-models-lqg}
\end{figure}

%=======================================================%
Here we present some additional spectra, particularly focusing on neutrino spectra (Figures~\ref{fig:neutrino_SuSy_ch} and \ref{fig:lqg_neutrino_SuSy_ch}). In Figures~\ref{fig:photon-particle-models-kerr} and \ref{fig:photon-particle-models-lqg} we compare across different values of the physical parameters describing the metrics: the spin values $a_*$ for the Kerr case and the polymerization parameter $\varepsilon$ for the LQG case.
%=======================================================%

%=======================================================%

%=======================================================%
%==========================================================

%=========================================================================
%======================================================================
\clearpage
\bibliographystyle{apsrev4-2}
\bibliography{bibliography}
%======================================================================

\end{document}